\documentclass[12pt]{iopart}
\usepackage{iopams}
\usepackage{graphicx}
\usepackage[usenames,dvipsnames]{xcolor}
\usepackage{url,cite}

\def\lsim{\mathop{\hbox{${\lower3.8pt\hbox{$<$}}\atop{\raise0.2pt\hbox{$\sim$}}
$}}} \def\gsim{\mathop{\hbox{${\lower3.8pt\hbox{$>$}}\atop{\raise0.2pt\hbox{$
\sim$}}$}}}  
\font\bigastfont=cmr8 scaled \magstep 3
\def\bolddot{\hbox{\bigastfont .}} 
 
\newcommand{\wlim}{\mathop{\mbox{w--lim}}} 
\newcommand{\liml}{{\lim_{\lambda \searrow 0}\,}} 
\newcommand{\CD}{{\cal D}}

\newcommand{\average}[1]{\left\langle #1 \right\rangle_\CD}

\newcommand\hMpc{\mbox{$h^{-1}$ Mpc}}

\newcommand\eprint[1]{\href{http://arXiv.org/abs/#1}{#1}} 

\providecommand\hMpc{\mbox{$h^{-1}$ Mpc}}

\usepackage{hyperref}
\hypersetup{
    colorlinks=true,
    citecolor=blue,
    linkcolor=magenta,
    filecolor=cyan,      
    urlcolor=cyan,
}
\begin{document}
\title[No proof that backreaction is irrelevant]{Is there proof that backreaction of inhomogeneities is irrelevant in cosmology?}
\author{T Buchert$^1$, M Carfora$^2$, G F R Ellis$^3$, E W Kolb$^{4}$, M A H MacCallum$^5$, 
J J Ostrowski$^{6,1, \dagger}$, S R\"as\"anen$^7$, B F Roukema$^{6,1, \dagger}$, \\L Andersson$^{8}$, A A Coley$^{9}$, and D L Wiltshire$^{10}$}
\smallskip
\address{ $^1$Universit\'e de Lyon, Observatoire de Lyon, 
Centre de Recherche Astrophysique de Lyon, CNRS UMR 5574: Universit\'e Lyon~1 and \'Ecole Normale Sup\'erieure de Lyon,
9 avenue Charles Andr\'e, F--69230 Saint--Genis--Laval, France \\
$^{2}$Dipartimento di Fisica, Universit\`a degli Studi di Pavia, via A. Bassi 6, I--27100 Pavia, Italy,
and Istituto Nazionale di Fisica Nucleare, Sezione di Pavia, via A. Bassi 6, I--27100 Pavia, Italy \\
$^{3}$Cosmology and Gravity Group, Department of Mathematics and Applied Mathematics, University of Cape Town, Rondebosch 7701, South Africa \\
$^{4}$Department of Astronomy and Astrophysics, Enrico Fermi Institute, The University of Chicago, Chicago, IL 60637, USA \\
$^{5}$School of Mathematical Sciences, Queen Mary University of London, London E1 4NS, UK \\
$^{6}$Toru\'n Centre for Astronomy, Faculty of Physics, Astronomy and Informatics, Grudziadzka 5, Nicolaus Copernicus University, ul. Gagarina 11, PL--87--100 Toru\'n, Poland \\
$^{7}$Department of Physics and Helsinki Institute of Physics, University of
Helsinki, FIN--00014 University of Helsinki, Finland \\
$^{8}$Max--Planck--Institute for Gravitational Physics (Albert Einstein Institute), Am M\"uhlenberg 1,
D--14476 Potsdam, Germany, and Department of Mathematics, Royal Institute of Technology, S--100 44 Stockholm, Sweden \\ 
$^{9}$Department of Mathematics and Statistics, Dalhousie University, Halifax, NS B3H 3J5, Canada \\
$^{10}$Department of Physics and Astronomy, University of Canterbury, Private Bag 4800, Christchurch 8140, New Zealand 
\\
\qquad\ $^{\dagger}$Lyon: BFR: during visiting lectureship; JJO: during long--term visit.}
%
\begin{abstract}
{\em No.} In a number of papers Green and Wald argue that the standard FLRW model approximates our Universe extremely well on all scales, except close to strong field astrophysical objects. 
In particular, they argue that the effect of inhomogeneities on average properties of the Universe (backreaction) is irrelevant. We show that this latter claim is not valid.
Specifically, we demonstrate, referring to their recent review paper, that 
(i) their two--dimensional example used to illustrate the fitting problem differs from the actual problem in important respects, and it assumes what is to be proven; (ii) the proof of the trace--free property of backreaction is unphysical and the theorem about it fails to be a mathematically general statement; (iii) the scheme that underlies the trace--free theorem does not involve averaging and therefore does not capture crucial non--local effects; (iv) their arguments are to a large extent coordinate--dependent, and (v) many of their criticisms of backreaction frameworks do not apply to the published definitions of these frameworks. 
It is therefore incorrect to infer that Green and Wald have proven a general result that addresses the essential physical questions of backreaction in cosmology. 
\end{abstract}
\vspace{-16pt}

\section{General context}

{\em Backreaction}, {\em i.e.}, the effect of inhomogeneities in matter and geometry on average cosmic evolution, has
been studied in the cosmology community from various perspectives (see the reviews \cite{buchert:review,chrisreview,ellis:focus,kolb:focus,buchert:focus,wiltshire:focus,rasanen:focus,buchertrasanen} and references therein).
A series of influential papers by Green and Wald \cite{gw1,gw2,gw3,gw} have led many to believe that these effects are irrelevant or highly constrained. 
Those papers are an interesting contribution to the research field of backreaction effects in relativistic cosmology, but in this paper we demonstrate that the strong claims advanced by Green and Wald about the irrelevance of backreaction are unproven.

In an earlier paper, Ishibashi and Wald \cite{ishibashiwald} claimed that backreaction effects are negligibly small on all scales, except ``in the immediate vicinity'' of strong field astrophysical objects. This statement was based on the smallness of metric perturbations, despite the presence of large density perturbations \cite{futamase1,futamase2}.  This conclusion evolved in subsequent papers \cite{gw1,gw2,gw3,gw}, where it was stated that backreaction effects can in principle be large (due to the fact that derivatives of metric perturbations are large \cite{estim}), but they can in fact only contribute a radiation--like
trace--free term to the effective stress--energy tensor. In this paper we will give evidence that backreaction effects can be significant and show that they need not be trace--free.

We organize this paper in line with the presentation of arguments in the recent
Green and Wald overview \cite{gw}, and refer by the abbreviation GW to the statements made therein.
In section \ref{sec:flatland} we consider the heuristic example of modelling
a polyhedron with a sphere discussed by Green and Wald in \cite{gw}.
We argue that this two--dimensional example misses essential features
of backreaction.
In section~\ref{sec:trace} we discuss the claim by Green and Wald that
the contribution of backreaction to the effective stress--energy tensor
is trace--free, and note that a non--trace--free part of backreaction is well--established;
here we also comment on problems with the weak--limit
procedure and its problematic relation to spatial averaging.
In section~\ref{sec:ex} we discuss some relevant examples.
In section~\ref{sec:scalaraveraging} we comment on some misinterpretations
related to scalar averaging, and in section~\ref{sec:obs} we put the discussion by Green and Wald of 
the relation to observations into perspective.
We conclude in section~\ref{sec:conc}. Appendices are devoted to technical aspects.

\section{An example of a fitting problem} 
\label{sec:flatland}

We start with the heuristic example in GW \cite{gw}. They state that
this solves a fitting problem. The fitting problem is raised and
explained in Refs. \cite{ellis:average,ellisstoeger,hellaby:volume}
and expressed in \cite{EllMaaMac} as ``How do we determine what is the
best FLRW background model for the real lumpy universe?''. In GW's
section 4.3 they argue for a parameter fitting approach in which it is
\emph{a priori} assumed that the best cosmological model is an FLRW model
with no backreaction. By making a similar assumption in their
introduction they conclude ``Thus, the `fitting problem' is trivially
solved''. From our perspective, it has not been satisfactorily
answered, in that GW have \emph{a priori} excluded the possibility that the
best FLRW model might be one with backreaction corrections. (The
real Universe might even be best modelled by some other geometry
altogether).

\subsection{Sphereland or Flatland?}

The GW example is of the physical spherical surface of a ``ball
bearing'' with all its ``tiny defects'', and an \emph{a priori}
assumed fit by the standard round geometry of the two--sphere. As a
model of the actual surface they take a polyhedral sphere
$(\mathbb{S}^2,\,g_{\mathrm pol})$ and consider how it is approximated
by the (model) metric of a standard round sphere
$(\mathbb{S}^2,\,g_{\mathrm can})$ (``can" for
  ``canonical"). The analogy they wish to draw is of an approximately
FLRW metric such that the difference of the real and model metrics
obeys the assumptions in their section 2. We discuss these assumptions
and their consequences at length in subsection \ref{subsection:scheme}
and in \ref{appendixB}.

Green and Wald assume $(\mathbb{S}^2,\,g_{pol})$ to be a convex
polyhedron. However, convexity is not an intrinsic characteristic of
$(\mathbb{S}^2,\,g_{pol})$. {\em It comes about only when we consider
  $(\mathbb{S}^2,\,g_{pol})$ embedded into Euclidean space}. The
assumption of convexity involves {\em extrinsic}
curvature, whereas the fitting problem is concerned with {\em
  intrinsic} curvature (as it must be in the cosmological setting).

Green and Wald discuss intrinsic aspects, by referring to E. Abbott's {\em
  Flatland} \cite{Flatland}. They argue that when two--dimensional
observers in `Sphereland' ``use triangles that are large compared with
the distance between vertices, they obtain results that are reasonably
consistent with spherical geometry. Nevertheless, even when looking at
phenomena on large scales, the observers find some disturbing
anomalies when attempting to model Sphereland by a perfect sphere''
and conclude that ``As a result of these observations, the observers
might be tempted to conclude that the perfect sphere model provides a
reasonably good description of Sphereland on large scales---although
with some significant deviations---but an extremely poor description of
Sphereland on small scales. However, the actual situation, of course,
is that the metric, $g_{ab}$, of Sphereland is everywhere extremely
close to the metric, $g_{ab}^{(0)}$, of a perfect sphere.''

The analogy is that cosmological observers may be similarly misled
into thinking an FLRW model (without backreaction) was a poor
description. This was made very clear in the recent note \cite{gwnote}
responding to the first draft of the present paper.

As quoted above, Green and Wald consider the Spherelanders performing geodesy with
larger and larger triangles, thus exploring the (global aspects of
the) intrinsic geometry of $(\mathbb{S}^2,\,g_{\mathrm pol})$. By this
means Spherelanders might, as GW suggest, reach the {\em apparently}
obvious conclusion that $g_{pol}$ is indeed approximated by the round
metric $g_{can}$, with some small--scale anomalies. However, the
measurements of the geometry of triangles proposed by GW do not
measure the metric, they measure curvature (integrated over the
interiors of the triangles). Knowing the curvature and a finite number
of its derivatives does locally specify a metric uniquely, but at the present time we do not know a way
to use this information to determine a best--fit approximate metric: we discuss
these points in subsection \ref{subsection:coorddep}.

More relevantly, it is the ``significant deviations'' on small scales
that are the important feature. Let us consider the steel ball example
in more detail.  The metric of each two--cell in the polyhedron
$(\mathbb{S}^2,\,g_{\mathrm pol})$ is {\em not curved}, but {\em
  flat}. Green and Wald recall that the ball's (Gaussian) curvature is
represented by a Dirac measure with support at the vertices of
$(\mathbb{S}^2,\,g_{pol})$, and geometrically represented by vertex
deficit angles measuring excess or defect with respect to the $2\pi$
vertex angles of a flat geometry. The latter implies, however, that
the geometry of $(\mathbb{S}^2,\,g_{\mathrm pol})$ is
\emph{almost--everywhere flat} with curvature defects supported at the
vertices: the relevant \emph{background metric} of
$(\mathbb{S}^2,\,g_{\mathrm pol})$ {\em is thus not a spherical
  metric}, as intuition and Figure 2 in GW's paper might suggest, {\em
  but a flat metric} (with conical singularities representing
curvature).

To illustrate what knowing the curvature, as the Spherelanders might,
means in the present setting, note that we may have positive
curvatures as well as negative curvatures ``sprinkled" over the
vertices of $(\mathbb{S}^2,\,g_{\mathrm pol})$. The crucial point is
that the ``sprinkling" {\em cannot be random}: the amount of overall
positive and overall negative curvature over the vertices must comply
with the Gauss--Bonnet theorem applied to $(\mathbb{S}^2,\,g_{pol})$
\cite{Troyanov}, \cite{Troyanov2}, \cite{Picard}; the integrated
Gaussian curvature must sum up to $4\pi$, the Euler number
$\chi(\mathbb{S}^2)\,=\,2$. Thus, any two--surface which is
topologically a sphere will have the same curvature, on average, as a
round sphere. Only by measuring over the whole surface would
Spherelanders find the correct total curvature, and anyway they would
arrive at the same result (assuming the same topology) regardless of
whether or not the coordinate components\footnote[1]{We pick up the issue
  of the use of coordinate components in comparisons in subsection \ref{subsection:coorddep}.} of
the metric were close, either in pointwise value, or in some other
suitable norm, to those of the round sphere.

\subsection{The consequences for the fitting and backreaction problems} 

To emphasize the point, we may easily construct a large (let it be
convex) polyhedral surface $(\mathbb{S}^2,\,g_{\mathrm pol})$, (using
a dual metrical triangulation), where thousands of vertices are flat
and just a few carry some (tiny) deficit angles.  For the Sphereland
inhabitants it would be very hard---perhaps impossible---to conclude
that they actually live on a two--sphere.  By probing rather large
portions of their ambient space (analogous, say, to the observable
Universe considered as part of an even larger spacetime) they would be
more likely to conclude, potentially with an arbitrarily high degree
of precision, that the geometry they live in is flat (there is no way
they could intrinsically detect the edge--bending between the flat
regions). Thus, guided by common sense and by such empirical evidence,
they would probably consider as irrelevant to their cosmological
modelling the local defects associated with a few deficit angles in an
ocean of flatness. However, it is these local defects that determine the
round sphere best--fit background, because it is the Euler number that
determines the global properties of a two--dimensional surface.  This
is the only datum which would allow Spherelanders to draw GW's desired
conclusions on the best--fitting metric in this analogy, and it is
thus the curvature defects that constitute the only relevant property.

We remark that in the mathematical literature this issue has been treated
in much greater depth than GW provide, giving more detailed presentations
of polyhedral manifolds and of the resulting approximation techniques, as indicated in \ref{appendixA}.

The steel ball analogy cannot say much about backreaction effects,
especially on their evolution, since the topological constraint makes
any two--dimensional example trivial.  Because of the Gauss--Bonnet
theorem, there is a conservation law for the integrated intrinsic
curvature in two dimensions \cite{magni,buchertrasanen}, just as there
is such a conservation law in the standard FLRW cosmological model in
$3+1$ dimensions.  As a consequence, if one imagined the
two--dimensional polyhedral model as representing the spatial surfaces
in a three-dimensional spacetime, the curvature evolution would be
decoupled from backreaction.

Generally, backreaction effects may become strong when inhomogeneities dynamically couple to the averaged intrinsic curvature (which does not, in general, admit a conservation law in three dimensions \cite{buchertcarfora:curvature}). This averaged curvature--backreaction coupling only arises in more than two dimensions (for details on the coupling of inhomogeneities to curvature and conservation laws, see \cite{buchert:review}).

\subsection{Conclusion} 

We have argued that the fitting problem cannot be solved by declaring,
dictated by common sense, that the best--fitting metric is $g_{can}$
with localized defects.
Considering curvature fluctuations as being unconstrained and {\em
  assuming} that they average out to zero is tautological since this
assumes that curvature defects globally contribute nothing and, thus,
{\em assumes} that the geometry  $(\mathbb{S}^2,\,g_{\mathrm can})$
{\em is} the best--fitting geometry without actually characterizing
the explicit map between  $(\mathbb{S}^2,\,g_{\mathrm pol})$ and
$(\mathbb{S}^2,\,g_{\mathrm can})$ (for discussion of such a map see \ref{appendixA}). Moreover, the geometry the Spherelanders experience in their surroundings (the physical geometry) is the almost--everywhere {\em flat} $g_{pol}$, not the everywhere {\em round} $g_{can}$. In particular, the measured physical curvature defects are with respect to  $(\mathbb{S}^2,\,g_{\mathrm pol})$ and not with respect to $(\mathbb{S}^2,\,g_{\mathrm can})$.

From this brief analysis we see why Green and Wald obtain a highly
constrained result for the possible backreaction in the steel ball model.

\section{Is backreaction necessarily trace--free ?}
\label{sec:trace}

{\em No}. There are many papers using standard perturbation theory and exact solutions of the Einstein equations that conclude that the backreaction has a trace part (see the reviews cited in the introduction and the references therein).  But Green and Wald claim that within a ``completely general framework'' \cite{gw} the backreaction is {\em trace--free}.  

\noindent
In this section we discuss the trace component in backreaction and show that GW's theorems are unrelated to actual backreaction.  Before doing so we look at the situation in the Newtonian limit.

\subsection{Traces of backreaction in Newtonian cosmology}
\label{subsect:newtonian}

The term {\em backreaction}, as it is usually understood in Newtonian or relativistic cosmology, may be generally defined as {\em deviation of spatial average properties of an inhomogeneous universe model from the values predicted by a homogeneous--isotropic universe model}.

\noindent
In Newtonian cosmology the average expansion rate is affected by the non--local term \cite{buchertehlers}
\begin{equation}
\label{q}
{\cal Q}_{\cal D} : = \frac{2}{3} \left( \average{\theta^2} - \average{\theta}^2 \right) - 2 \average{\sigma^2} + 2 \average{\omega^2}  \;,
\end{equation}
where $\theta$, $\sigma$ and $\omega$ are the rate of expansion, shear and vorticity for a given velocity model, respectively, and the brackets denote volume averaging on a spatial domain $\cal D$, an averaging that is well--defined even for tensors in Newtonian theory. 
The magnitude of this term has been estimated in \cite{bks} (and follow--up papers) using well--known models for large--scale structure that are in good agreement with $N-$body simulations down to scales where the dust approximation breaks down. 

Interpreted in the context of a FLRW fitting model, backreaction leads to an
{\em effective} stress--energy tensor that has the form of a perfect fluid with the components (see, e.g. \cite{buchert:morphon}, Subsection 3.2):
\begin{eqnarray}
\label{Tnewton}
\varrho_{\rm eff} : &= \average{\varrho} + \frac{1}{8\pi G a_{\cal D}^2} \int_{t_i}^{t} {\cal Q}_{\cal D}\frac{d}{dt'} a_{\cal D}^2 (t') \; {\mathrm d}t'   \quad;\quad\nonumber \\
p_{\rm eff}: &= - \frac{1}{24\pi G a_{\cal D}^2} \int_{t_i}^{t} {\cal Q}_{\cal D}\frac{d}{dt'} a_{\cal D}^2 (t') \; {\mathrm d}t' - \frac{1}{12\pi G} {\cal Q}_{\cal D} \;,
\end{eqnarray}
where $a_{\cal D}$ is the volume scale factor, defined such that the volume of the
averaging domain is proportional to $a_{\cal D}^3$ \cite{buchert:dust}. 
The {\em local} stress--energy tensor satisfies the weak energy condition, since $\varrho \ge 0$ (implying $\varrho + p \ge 0$ as we only consider dust, with $p=0$).
However, the {\em effective} stress--energy tensor does not necessarily
satisfy these conditions, and neither does its backreaction part\footnote[2]{This latter condition would artificially select the class of models for which $\varrho_{\rm eff} - \average{\varrho} \ge 0$ and $\varrho_{\rm eff} - \average{\varrho} + p_{\rm eff} \ge 0$ hold.}. Also, the trace is not in general zero.
Both of these features are in contradiction with GW's claims, specifically their Theorems 1 and 2 \cite{gw1,gw}, considered in the Newtonian limit.
We shall come back to the energy conditions in subsection~\ref{subsect:energycondition}.

Note that the magnitude of backreaction depends on the scale of averaging.
In particular, already in Newtonian cosmology it is significant on intermediate
scales, below the cosmological homogeneity scale\footnote[3]{E.g., in a domain with a radius of $100$Mpc today and initially one--$\sigma$ fluctuations, the density parameters deviate from their homogeneous values of an Einstein--de Sitter background by $15\%$\cite{bks}; for estimates in relativistic cosmology including a background with cosmological constant see \cite{rza2}).}, but well above the scales of ``black holes or neutron stars'',
excluded by Green and Wald\footnote[4]{We note that there are two cases in Newtonian cosmology in which this backreaction term vanishes, both being the result of the flat geometry in Newtonian theory: first, it vanishes  for spherically symmetric solutions (the content of Newton's ``theorem of the iron spheres'') \cite{bks} and, second, for scales where commonly periodic boundary conditions (a $3-$torus topology) are imposed on the deviations off a homogeneous solution (a topological constraint as in the steel ball model) \cite{buchertehlers}. Thus, backreaction vanishes globally, but not in the interior of this simulation box where it is significant on smaller scales. However, if we additionally perform a statistical average over many realisations (not considered by GW), this term vanishes in the statistical average due to the global constraint \cite{bks}.}. 

\vspace{-5pt}

\subsection{Traces of backreaction in relativistic cosmology}
\label{subsect:GR}

Since the trace part of backreaction cannot turn on just in the Newtonian limit, backreaction cannot be trace--free {\em in general} in GR either without violation of at least one of the theory's principles.  This remark makes the existence of trace parts clear, since the calculation in the Newtonian framework is unambiguous. Green and Wald emphasize in their abstract that ``Newtonian cosmologies provide excellent approximations to cosmological solutions to the Einstein equations (with dust and a cosmological constant) on all scales.''
If that is the case, then---especially in view of their ``dictionary'' \cite{gw2}---a claim of trace--free backreaction cannot be correct.

General Relativity adds the new feature that this backreaction term (represented by extrinsic curvature invariants here) is coupled to the averaged scalar curvature $\average{\cal R}$ \cite{buchert:dust},
\begin{equation}
\frac{1}{a_{\cal D}^6} \left( {\cal Q}_{\cal D} a_{\cal D}^6 \right){\dot{}} + \frac{1}{a_{\cal D}^2} \left( \average{\cal R}a_{\cal D}^2 \right){\dot{}}= 0\;\;.
\end{equation}
This coupling is important for backreaction in relativistic cosmology, as it can also affect the global geometrical properties. This latter is impossible in Newtonian cosmology and it is also suppressed in quasi--Newtonian relativistic perturbation theory with periodic boundary conditions due to the fact that a conservation law for the intrinsic curvature holds.

\vspace{-8pt}

\subsection{The idea of Green and Wald and its realisation}

Green and Wald assume that the actual metric has the structure:
\begin{equation}
\label{one}
g_{ab}\,=\,g_{ab} (0)\,+\,\gamma_{ab}\;,
\end{equation}
where $g (0)$ (denoted by $g^{(0)}$ in GW) is a background metric assumed to provide the averaged background modelling the Universe (the indices are spacetime indices). Green and Wald stress 
that in most of their computations $g (0)$ can be any given background metric.

Carrying out computations in such a general framework (and in particular trying to interpret them geometrically) is a daunting, if not impossible, task. We are  not aware of any similar analysis in Riemannian geometry (where things are definitively easier than in the Lorentzian case).
Hence, Green and Wald tackle a less ambitious task by considering a
perturbative formalism where the symmetric tensor $\gamma$, representing the deviations
from the background, is small. However, they do not assume that the
first derivatives of $\gamma$  are small and even allow second derivatives of $\gamma$ to be unboundedly large, to
deal with very large fluctuations in curvature around the $g(0)$ background.

We point out already here, with detailed explanations in \ref{appendixB}, that the assumption in Equation (\ref{one}) 
together with the non--perturbative implications for the metric derivatives outlined above,
inherits similar problems to those we explained for the steel ball analogy. Their statement 
that ``In summary, the geometry of Sphereland is described by a metric of the form $q_{ab}(0) 
+ s_{ab}$, where $q_{ab} (0)$ is the metric of a perfect sphere and $| s_{ab}| \ll |q_{ab}(0)|$, but
first derivatives of $s_{ab}$ are not small, and second derivatives of $s_{ab}$ may be enormous.
In an exactly similar manner, the {\em spacetime} metric of
our Universe takes the form $g_{ab}=g_{ab}(0)+\gamma_{ab}$, where  $
g_{ab} (0)$ has the FLRW symmetry ..." \cite{gw} is problematic. Modelling the perturbation $\gamma_{ab}$  with the same underlying
rationale as that for the perturbation $s_{ab}$ of the steel ball example
implies that the potentially unbounded second derivatives of the metric
may give rise to curvature distributional defects in the weak--limit, {\em cf.} \ref{appendixB}, that cannot be dismissed, since they are as crucial as the curvature defects in the steel ball example.

The scheme adopted by GW adds a stress--energy tensor to a scheme developed by Burnett \cite{burnett} for vacuum spacetimes, which replaces the averaging operation in Isaacson's approach \cite{isaacson1,isaacson2} by a weak--limit procedure. (Burnett \cite{burnett} finds a trace--free effective stress--energy tensor by applying a weak--limit scheme to gravitational waves, which are physically trace--free. He also discussed a non--vanishing stress--energy--tensor, but for the case of electromagnetic radiation, which is again physically trace--free.) For another calculation of backreaction in the weak--limit of gravitational waves emitted from a Schwarzschild black hole see  Ref. \cite{futamasehogan}. We shall demonstrate that the weak--limit operation does not implement a true averaging procedure, and that the assumptions made by GW are too restrictive to allow for backreaction.

\vspace{-5pt}

\subsection{The weak--limit scheme}
\label{subsection:scheme}

Green and Wald's weak--limit scheme is technically realised, as in the familiar perturbation formalism for GR, by assuming that there is a class of coordinate systems $\{x^a\}$ and a one--parameter family of Lorentzian metrics $(0,1]\,\ni\,\lambda\,\longmapsto \, g(\lambda)$, such that we can locally write the metric components $g_{ab}(\lambda)$ as  
\begin{equation}
\label{two}
(0,1]\,\ni\,\lambda\,\longmapsto \, g_{ab}(\lambda)\,=\,g_{ab} (0)\,+\,\gamma_{ab}(\lambda)\;.
\end{equation}
These metric components are further required to satisfy the following assumptions: ({\sl cf.} $(i)$--$(iv)$ of GW):\\
\noindent $(i)$\; For all $\lambda\in (0,1]$  one assumes that the Einstein equations hold,
\begin{equation}
\label{three}
G_{ab}(g(\lambda))\,+\,\Lambda \,g_{ab}(\lambda)\,=\,8\pi\,T_{ab}(\lambda)\;,
\end{equation}
where $T_{ab}(\lambda)$ is a one--parameter family of stress--energy tensors that locally satisfy the weak energy condition.\\
\noindent $(ii)$\;There exists a smooth positive function $C_1(x)$  on $(M, g (0))$ such that we may 
write\footnote[5]{Green and Wald use a fixed $\lambda$--independent Riemannian metric for computing norms of the tensor field involved and of their covariant derivatives, {\sl cf}. Equations (5) and (6) of GW.}
\begin{equation}
\label{four}
\left| \gamma_{ab}(x, \lambda)  \right|\,\leq\,\lambda\,C_1(x)\;.
\end{equation}
This is the usual way of controlling the \emph{perturbing} tensor $\gamma$ in standard perturbation theory, implying that,\, as $\lambda\searrow 0$, \, $g(\lambda)$ approaches $g (0)$.\\
\noindent $(iii)$\;There exists a smooth positive function $C_2(x)$ on $(M, g (0))$ such that we may write
\begin{equation}
\label{five}
\left|\nabla _c \gamma_{ab}(x, \lambda)  \right|\,\leq\,C_2(x)\;,
\end{equation}
where $\nabla $ denotes the covariant derivative w.r.t. $(M,\,g (0))$. This requirement does not imply that the first derivatives of $\gamma$ are small
and they do not necessarily become small as $\lambda \searrow 0$. \\
\noindent $(iv)$\;Finally, they assume that there exists a smooth tensor field  $\mu_{abcdef}(x)$  on $(M,\,g (0))$  representing the weak--limit of the product of first covariant derivatives  $\nabla \gamma(\lambda)\nabla \gamma(\lambda) $ as $\lambda\searrow 0$,\,\emph{i.e.},
\begin{equation}
\label{six}
\wlim_{\lambda\searrow 0}\,\left| \nabla _a \gamma_{cd}(x, \lambda)\nabla _b \gamma_{ef}(x, \lambda)\right|\,=\,
\mu_{abcdef}(x)\;.
\end{equation}
Here, the weak--limit is intended in a strict weak sense, \emph{i.e.}, requiring that the functions $\mu_{abcdef}$ are locally summable (see GW, Eqs. (5) and (6)). 
(It should be noted that this is more restrictive than the usual distributional interpretation of the weak--limit, where $\mu_{abcdef}$ could be represented by a singular distribution. We examine this issue in depth in \ref{appendixB}). This condition forces good behaviour of the potentially problematic quadratic product $\nabla \gamma(\lambda)\nabla \gamma(\lambda) $ as $\lambda\searrow 0$.

The condition $(iv)$ is not motivated by any  transparent and natural
physical argument: as it stands we do not consider $(iv)$ to be a robust hypothesis
underlying backreaction analysis. Even for the high--frequency limit regime of gravitational waves
there are physically transparent alternative approaches to this that do not need to assume condition $(iv)$, see \cite{Choquet},
Ch.~III, paragraph 12 and Ch.~XI, (this reference is also interesting for the issue of gauge dependence, discussed further below), and also \cite{Mad72,MacTau73,Ara86}
and references therein.

\vspace{-5pt}

\subsection{Green and Wald's one--parameter family of metrics and metric flows}

A delicate issue with GW's perturbation approach is the validity of assumption $(iv)$ together with assumption $(i)$ that {\em the Einstein equations are required to be satisfied along the curve of metrics, for all $\lambda$ except in the limit $\lambda\searrow 0$}. Thus, backreaction is supposed to turn on just in the limit (see subsection~\ref{nobackreaction} for details). This situation will turn out to be problematic. 

We first point out an issue that arises in taking due care of the subtleties generated by the potentially unbounded second derivatives of the metric fluctuations.
There may be particular spacetimes for which their procedure can be carried out, but typically one would get {\em distributional curvature tensors} under the GW hypotheses in the limit where $\lambda\searrow 0$, which makes their effective stress tensor $t_{ab} (0)$, {\em cf.} subsection \ref{nobackreaction}, just a formal expression, {\em not containing functions but distributions}, which are not easily interpreted from the geometrical and physical point of view. In particular the trace--free conditions would generally be false. Even it were possible to get rid of any non--regular distributional part, as GW implicitly assume, 
their approach  strongly relies on a very delicate non--uniform boundedness (in $\lambda$) hypothesis on $\left|\nabla _c \nabla_d\gamma_{ab}(x, \lambda)  \right|$ (see the discussion surrounding Eq.~(\ref{tricky})).  This feature is imported from Burnett's high--frequency limit in gravitational wave theory, and it is precisely this non--uniformity that generates a backreaction only in the $\lambda\searrow 0$ limit. However, whereas in gravitational wave theory a non--uniform 
high--frequency limit is not a surprising feature of oscillating phenomena,  in cosmology it 
does not correspond to the physical situation. In the real Universe, the
matter distribution does not oscillate on arbitrarily small scales, but there is
a hierarchy of finite averaging length scales that are physically defined by the gravitational systems we want to average. It is important to stress that if we remove this specific
non--uniformity requirement from the GW formalism, for example instead requiring the second derivatives of $\gamma$ to be uniformly bounded in $\lambda$,  then
{\em the effective stress--energy tensor $t_{ab}(0)$ vanishes.} We give the proof in \ref{appendixB}.
Moreover, in a real averaging procedure, {\em backreaction terms must already be present for non--zero $\lambda$}.
We shall explain this remark in the following two subsections.

We also note that known and well--controlled metric flows as investigated in the mathematical literature on cosmological backreaction are generally not required to satisfy the Einstein equations during metric deformations. Metric flows are typically (weakly) parabolic and, hence, in general they do not commute with the Einstein evolution which is (weakly) hyperbolic with elliptic constraints; `weakly' here refers to the fact that these flows must comply with the appropriate form of  equivariance under the action of the diffeomorphism group (see Ref. \cite{fischermarsden}). Geometric flows are often gradient flows (with respect to some Hilbertian positive definite inner product), whereas the Einstein evolution is Hamiltonian.  In particular, in a cosmological setting, rescaling flows (such as the Ricci--Hamilton--Perelman flow) work in a 3D hypersurface, rescaling the metric in the direction of the spatial Ricci tensor (see Ref. \cite{carfora}), which {\em a priori} is unrelated to the Einstein flow that deforms the initial data set along the (symplectically conjugated) linearization of the Einstein constraints in the direction of the lapse and shift. In other words, the former relate to flows in space and the latter to time evolution.

\vspace{-5pt}

\subsection{Averaging without averaging?}
\label{noaverage}

While Green and Wald aimed at formulating the weak--limit formalism in a mathematically ``completely general'' manner, they noted that ``it would be extremely difficult to formulate mathematically precise criteria for the validity of applying our formalism to a given spacetime''. They argued that they simply ``resort to plausibility arguments to obtain [their] conclusions'' regarding the applicability of the formalism to the real Universe \cite{gw1}.
They then argued that it is plausible that the weak--limit corresponds to spacetime averages over local regions, provided that the size of the regions corresponds to the homogeneity scale, which is much smaller than the curvature scale of the background universe. However, arguments about limits that may be plausible for smooth functions do not immediately translate to singular distributions.

Formally, GW are {\em not} averaging the fluctuations represented by the tensor $\gamma$. They  consider the weak--limit in a \emph{singular perturbation scheme} for GR around a FLRW spacetime (or whatever geometry $g (0)$ represents). In their words (GW): ``The notion of `weak--limit', which appears in assumption (\emph{iv}), corresponds roughly to taking a local 
spacetime average, and then taking the limit as $\lambda\searrow 0$". This is not averaging in the sense used in discussing backreaction.
The weak--limit of a sequence of  smooth functions is typically a (singular) distribution. An example (on a compact Riemannian manifold) is provided by the heat kernel: $K(x,y;t)$,\, $t>0$. This is $C^\infty$ for all $t>0$, so smooth that its convolution against nasty functions tames all their defects.  Nonetheless the weak--limit of  $K(x,y;t)$,  as $t\searrow 0$, is a singular distribution, the Dirac measure $\delta_y$ supported at the point $y$.
The operation might correspond to averaging if the compactly supported function that GW used to define the weak--limit had a form of translation symmetry built in  (say, $\phi(x-y)$ in $\mathbb{R}^n$, or if we consider a large symmetry group, as in FLRW), so that one is actually \emph{regularising} (the distribution) via a convolution product, not just a generic weak--limit. 

Green and Wald consider a singular perturbation theory around FLRW, {\em not} averaging or taking convolutions. This strategy is quite common in quantum field theory (QFT) where one needs to tame the spectrum of fluctuations of a quantum field around a given background configuration and describe how these fluctuations may dress the bare background\footnote[6]{For the difference between dressed and bare backgrounds in cosmology see Ref.\ \cite{dressing}.}.

Dismissing the potential distributional nature of curvature fluctuations in such a setting is akin to the situation we encountered in the steel ball example. The analogy is (i) here the background metric $(M,\,g (0))$ plays the role of the almost--everywhere flat background metric in the steel
ball model; and (ii) the distributional part 
plays the role of the physical curvature fluctuations described by the conical vertices in the steel ball model. As in that case, there is no reason why they should disappear when performing a real averaging procedure. (In the two--dimensional steel ball model the Gauss--Bonnet theorem implies that they cannot disappear.)

Clearly, there is nothing wrong in addressing the backreaction issue by studying a singular perturbation scheme around FLRW as GW do. It is a potentially useful idea. After all, in relativistic cosmology, unlike the averaging of scalars (which we will come to below), {\em the averaging of tensors} is not yet well--established, at least in a rigorous sense. The real point is to understand what you can get out of such an approach, how it is technically realised, and how it can be physically interpreted.

We finally remark that, even if not gauge invariant, a singular limit may be given physical sense {\em locally}, e.g., where Newtonian spacetime is obtained as a singular limit of a sequence of Einstein spacetimes (defined by appropriate initial data expressing the Newtonian
situation so that it can only be defined near the limit) \cite{futamaseschutz}.

\subsection{Backreaction from no backreaction?}
\label{nobackreaction}

Even if one were to extend the GW perturbation scheme to rigorously calculate the trace from the distributional curvature using the procedure outlined in \ref{appendixB}, there is still another issue: Green and Wald assume the Einstein equations to hold for all $\lambda$, {\em except for} $\lambda = 0$. This is shown in their assumption $(i)$, recall Equation~(\ref{three}):
$$
G_{ab}(g(\lambda))\,+\,\Lambda \,g_{ab}(\lambda)\,=\,8\pi\,T_{ab}(\lambda)\quad;\quad \lambda \in (0, 1] \;,
$$
and in their terminology for the limit (GW, Equation (1)):
$$
G_{ab}(g(0))\,+\,\Lambda \,g_{ab}(0)\,=\,8\pi \left( T_{ab}(0) + t_{ab} (0) \right) \;.
$$
Green and Wald refer to the averaging and fitting problems as they were raised in Ellis \cite{ellis:average} and Ellis and Stoeger \cite{ellisstoeger} (GW's Figure 1 is reproduced from \cite{ellis:average}). Those references emphasise that 
averaging or smoothing the inhomogeneous metric and taking derivatives (and their products) are in general non--commuting operations, {\em which implies that, in general, the Einstein equations do not hold at any smoothing level}$\,$\footnote[7]{See \cite{ellisbuchert} for a summary and examples of non--commuting operations.}.
In other words, although the actual model obeys the Einstein equations, there need not
be a one--parameter family of solutions approaching the background that also obey the Einstein equations.
This issue lies at the basis of backreaction, and the first--principle argument of \cite{ellis:average} can be realised, e.g., with the scalar averaging formalism discussed in Section~\ref{sec:scalaraveraging},  or (directly related to a smoothing procedure) with Ricci flow techniques \cite{klingon}.
 
Putting this argument formally and adopting GW's notation, an averaging operation, however it is defined (and here specialized to a one--parameter family of metrics), leads to effective terms, symbolically put into the tensor $\tau_{ab}$\footnote[8]{This term is interpreted as an effective stress--energy tensor, although it arises from smoothing the geometrical side.}:
\begin{equation}
\label{ellisargument}
G_{ab}( g(\lambda)) \,+\,\Lambda \,g_{ab}(\lambda)\,=\, 8\pi \left( T_{ab}(\lambda) + \tau_{ab} (\lambda) \right)\quad;\quad\lambda \in (0, 1) \;.
\end{equation}
Green and Wald correctly quote the starting point of Ellis' argument \cite{ellis:average} that the Einstein equations are assumed to hold in the unaveraged case ($\lambda =1$), but they then assume that the Einstein equations also hold for $\lambda < 1$ down to the limit, where suddenly the backreaction term $t_{ab}$ 
turns on\footnote[9]{In their recent note in response to the draft of this paper \cite{gwnote} Green and Wald claim that ``backreaction in our [GW's] formalism does not suddenly `turn on' at $\lambda = 0$''. This is in disagreement with the requirement that the Einstein equations are assumed to hold for all $\lambda \ne 0$ and the fact explained in the previous subsection that no averaging is involved. The statement that ``... the dynamics of  $g_{ab}(0)$  accurately describes the (large
scale) dynamics of $g_{ab} (\lambda)$ for sufficiently small $\lambda$" does not take into account the lack of continuity of the mapping $g_{ab}(\lambda) \mapsto G_{ab}(\lambda)$, as we explain in \ref{appendixB}. Moreover, the claim that ``Éfor $\lambda > 0$, backreaction terms are
present, and are described to leading order by the second order Einstein
tensor'' is  not  to do with backreaction arising from averaging, but the difference $G_{ab}(\lambda)- G_{ab}(0)$ (see Eqs. (9)--(12) in \cite{gw1}). If such terms were backreaction terms, then in gravitational perturbation theory every term of order higher than $0$ should be
considered as  a backreaction term for the assumed background.}.

In the framework of the weak--limit scheme, starting from (\ref{ellisargument}), the calculation of the weak--limit gives:
\begin{equation}
G_{ab}( g(0)) \,+\,\Lambda \,g(0)\,=\, 8\pi \left( T_{ab}(0) + t_{ab} (0) + \wlim_{\lambda\searrow 0}\; [ \tau_{ab}(\lambda)] \right)\;,
\end{equation}
where the last term is the weak--limit of the $\lambda-$dependent backreaction term, which is absent in GW's calculation.
We will give an argument in subsection~\ref{pathdependence} and a proof in \ref{appendixB} that the effective stress--energy tensor $t_{ab}(0)$, the trace--free ``backreaction term'' of GW,
is expected to vanish for a uniform convergence to the background metric.

We now illustrate the above issues in terms of examples. 

\section{Examples using the GW framework}
\label{sec:ex}

Green and Wald provide an example of a family of vacuum spacetimes in \cite{gw3} (section 3), discussed in 
subsection~\ref{pathdependence} below.  In this and our other examples we require, as before, $g_{ab} (x, \lambda)$ to be a solution to the Einstein equations for 
all $\lambda > 0$ (GW's assumption $(i)$), and,
as $\lambda \searrow 0$, we require convergence of $g_{ab} (x, \lambda)$ to a background metric $g_{ab} (x, 0)$.

We refer the reader also to \ref{appendixC}, where we comment on the example provided by  Szybka et al. \cite{szybka} and on another example by GW that they provided in \cite{gw3} (section 4), both of which aim at including matter inhomogeneities. 
We are not aware of any example satisfying the GW conditions that does satisfactorily include matter inhomogeneities.

We first start with a classical example given by Geroch to demonstrate the coordinate--dependence of the limit of a one--parameter family of metrics.

\subsection{Coordinate--dependence of the GW framework: Geroch's example}
\label{subsection:geroch}

It was emphasized by Geroch \cite{geroch} that, if we restrict our attention to 
metrics parametrised by some $\lambda$ only, then the limit is not uniquely defined.

Geroch considers the family of parametrised Schwarzschild metrics: 
\begin{equation}
\label{e-Geroch-Schw-family} 
{\mathrm d}s^2 = \left(1 - \frac{2}{\lambda^3 r}\right){\mathrm d}t^2 - \left(1 - \frac{2}{\lambda^3r} \right)^{-1} {\mathrm d}r^2 - r^2({\mathrm d}\theta^2 + \sin^2\theta {\mathrm d} \phi^2)\;,
\end{equation}
where $\lambda = m^{-1/3}$. In this form there is no limit as $\lambda \searrow 0$. Consider, however, the coordinate transformation: 
$\tilde r = \lambda r ,\;\; \tilde t = \lambda^{-1}t ,\;\; \tilde \rho = \lambda^{-1}\theta$.
Then, our family of metrics takes the form:
\begin{equation}
{\mathrm d}s^2 = \left(\lambda^2 - \frac{2}{\tilde r}\right){\mathrm d} {\tilde t}^2 - \left(\lambda^2 - 
\frac{2}{\tilde r} \right)^{-1}{\mathrm d} {\tilde r}^2 - {\tilde r}^2({\mathrm d} {\tilde \rho}^2 + \lambda^{-2}\sin^2(\lambda \tilde \rho){\mathrm d} {\phi^2})\;,
\end{equation}
and the limit exists -- it is  the Kasner metric with Kasner exponents $-1/3$, $2/3,$ $2/3$:
\begin{equation}
{\mathrm d}s^2 = - \frac{2}{\tilde r}{\mathrm d} {\tilde t }^2 + \frac{\tilde r}{2}{\mathrm d} {\tilde r}^2
- {\tilde r}^2 ({\mathrm d} {\tilde \rho}^2 + {\tilde \rho}^2 {\mathrm d} \phi^2)\;.
\end{equation}
However, applying a different coordinate transformation, $x = r + \lambda_0\lambda^{-4} ,\;\; \rho = \lambda_0\lambda^{-4}\theta$ (for an arbitrary constant
$\lambda_0$) to our original metric (\ref{e-Geroch-Schw-family}) and taking the limit $\lambda \searrow 0$ yields the flat Minkowski metric.  

Geroch's discussion illustrates that the outcome of the weak--limit procedure depends on the choice of coordinates. Moreover, it can be argued that instead of fixing the background metric, it should emerge from a {\em coordinate--independent} averaging or rescaling procedure.
We finally remark that the characterization by curvature invariants can be used to
clarify Geroch's example and avoid the coordinate dependence of his methods (see Ref. \cite{paiva}). Such ideas could work whenever there is some identifiable way of taking a weak--limit of the set of curvature invariants.

The issue of gauge dependence has been addressed in Ref. \cite{ellis:focus}. A related discussion can be found in \cite{veneziano1}. In this latter paper it is demonstrated that even the spacetime integrals of 4D scalars are in general gauge--dependent if the domain of integration is fixed. The domain should rather be defined intrinsically: the behaviour of the domain under gauge transformations should compensate the effects of the same gauge transformations applied to the integrated object and, thus, the domain must be coordinate-- and $\lambda$--dependent (the latter requirement comes from the parametrisation of the metric and its volume element). Given such an $x$-- and 
$\lambda$--dependent domain, it would no longer be allowed to exchange the limits and the integration operators as is done by GW in the case of first derivatives of metric deviations. This will then imply a non--commutativity of the two operations and would give rise to backreaction terms (for the consequences of non--commutativity see Ref. \cite{buchert:dust} for the Einstein flow, Ref. \cite{klingon} for three--dimensional metric flows, and Ref. \cite{ellisbuchert} for a general discussion). Then, the weak--limit operator would match the authors' expectations of first performing an average and then taking a limit. This $\lambda-$dependence has also been emphasized in a recent paper by Kopeikin and Petrov \cite{kopeikinpetrov}. 

We move now to another issue. As pointed out by Szybka \cite{szybka}, 
one of the key ingredients of the GW formalism is Burnett's
restriction that any $\lambda$--dependent coordinate transformation 
must be reduced to the identity when $\lambda \searrow 0$. This is essential and shows that the procedure cannot work for arbitrary $\lambda-$dependent coordinate transformations and is therefore not covariant. The Geroch coordinate transformations would be excluded with this condition. However, even if we take such ``allowed'' coordinate transformations we have a path dependence in the weak--limit procedure as discussed in the next subsection.

\subsection{Path--dependence of the weak--limit procedure}
\label{pathdependence}

In Ref.\ \cite{gw3} Green and Wald look at a subclass of vacuum Gowdy metrics on a torus with metric
\begin{equation}
{\mathrm d}s^2 = e^{(\tau - \alpha)/2} \left(-e^{-2\tau} {\mathrm d}\tau^2 + {\mathrm d}\vartheta^2 \right) + 
e^{-\tau}\left( e^P {\mathrm d}\sigma^2  + e^{-P} {\mathrm d}\delta^2 \right) \;,
\end{equation}
which depends on the function $P (\tau , \vartheta )$. A one--parameter family of metrics parametrised by a 
discrete parameter $N$\footnote[10]{The restriction to integer $N$ is related to a topological assumption, but
the solutions are still solutions if $N$ is replaced by a continuous parameter $1 / \lambda$, although they cannot then be 3D
spatially toroidal.} is {\em chosen} as follows:
\begin{equation}
\label{path}
P_N = \frac{A}{\sqrt N}J_{0}(Ne^{-\tau})\sin(N\vartheta) \;,
\end{equation}
so that, like $\lambda \sin(x/\lambda)$, for larger and larger $N$ (corresponding 
to smaller and smaller $\lambda$) it oscillates with shorter and shorter 
wavelength and smaller and smaller amplitude; the limit $N \rightarrow \infty$ corresponds to $\lambda \searrow 0$.

The weak--limit gives a non--zero result only for functions such as $f (x) = x \sin(1/x)$ (as 
in example (4.2) of \cite{gw3}), which have a singular limit as they oscillate ever more finely without
limit as $x \rightarrow 0$. Any non--zero result depends {\em essentially} on taking
the infinite limit of those oscillations as the homogeneous model is approached;
it will be zero if they cease at any value of $\lambda$, no matter how small. This is what is called in \cite{ellis:focus}  an {\em ultra--local limit}: 
the oscillations must not even stop at atomic scales.

But the path chosen to link $g_{ab} (x, N=1 )$ to $g_{ab} (x, 0)$ is arbitrary: there is no reason why the specific path of Eq.\ (\ref{path}) is preferred over others. One might equally well choose
\begin{equation}
P_N = \frac{A}{N}J_{0}(Ne^{-\tau})\sin(N\vartheta) \quad {\rm or} \quad
P_N = \frac{A}{N^2}J_{0}(Ne^{-\tau})\sin(N\vartheta)
\end{equation}
for example; these are both solutions of the field equations, 
{\em and these paths 
will give $t_{ab}(0) = 0$}. The physical entity is the endpoint $g_{ab} (x, N=1)$; the intervening family is arbitrary, 
as is the parameter choice. 

The result obtained from Eq.\ (\ref{path}) only has physical meaning if it gives the same 
answer for different such parametrisations. This procedure does not.

One could try
\begin{equation}
P_N = \frac{A}{f(N)}J_{0}(Ne^{-\tau})\sin(N\vartheta)\;,
\end{equation}
and see for what $f (N)$ one gets $t_{ab}(0) \neq 0$ as well as the amplitude $
A/f (N )$ going to zero. It is non--zero only for very special choices of $f (N )$.

We repeat that the problem is the ultra--local 
requirement that perturbations in the envisaged family
continue down to indefinitely small wavelengths without limit; otherwise one gets a zero answer.
For very small distances in realistic cosmological modelling
the solutions should behave like $(A/N )$ rather than like $(A/N ) \sin(N\vartheta)$ (equivalent to using $f (x) = x$ 
for very small scales as $x \rightarrow 0$ rather than $x \sin(1/x)$). Then, 
the quantity $t_{ab} (0)$ will be zero. Thus, Equations (2.7) and (2.8) of \cite{gw3} are true in realistic situations only because $t_{ab}(0)=0$. 

The pathological nature of the limit is emphasized by the fact that in
the example given, for each $N < \infty$, the metric is a solution of the vacuum field equations,
but the limiting metric (3.11) of \cite{gw3} is not a solution. The backreaction term
turns on in a delta--function way, in the exact limit only. Genuine backreaction
effects should not have a delta function discontinuity of this sort: the 
backreaction term should be non--zero for every metric where $\lambda \ge 0$ in any
realistic sequence approaching the background, as we have already pointed out.
In the real Universe, inhomogeneities have a finite amplitude and length scale,
which should correspond to some $\lambda = \lambda_0>0$. One could use a Taylor
series to estimate the effect at $\lambda = \lambda_0$ from the data at $\lambda = 0$,
but that will not work for this singular family of perturbations.

These examples demonstrate that $t_{ab}(0)$ should vanish for a realistic sequence of cosmological perturbations, which should be cut off at a {\em finite} wave--length before the amplitude goes to zero, inter alia because the Sun exists; real perturbations do not continue to zero wavelength. Indeed, as we show in \ref{appendixB}, the quantity $t_{ab}(0) $ is only non--vanishing if we impose further restrictions beyond the conditions imposed by GW and, especially, if we require non--uniform convergence so that this term only arises in the singular limit (see the discussion surrounding Eq.~(\ref{tricky})). 
It is worth noting that any uniformly convergent limit of a family of spacetimes will inherit some of the properties of the family. These properties are called
{\em hereditary} \cite{geroch}. A set of vacuum solutions with a non--vacuum limit would not respect Geroch's hereditary properties for limits, reflecting
the points made above about the GW scheme being singular.

\subsection{An example by Korzy\'nski}

We will not list here the many investigations of models and exact solutions illustrating or quantifying backreaction effects. These works may be found in the reviews mentioned in the introduction. We will only point out one recent work by Korzy\'nski on nonlinear effects from multi--scale structure \cite{korzynski} because it closely follows the arguments of GW, it precisely illustrates why their arguments do not apply, and it quantifies the GR inhomogeneity effect with the help of an exact nested structure model.

Korzy\'nski shows that it is possible for small local deviations to produce a
large global effect. The procedure is closely related to the idea of metric
smoothing by removing density ripples. His result is not a counter--example to
the claimed results of GW, because, as he states, his model violates
GW's assumption $\emph{(ii)}$, according to which perturbations should be small
not just locally, but with reference to a single global background.
In Korzy\'nski's example, the smoothing procedure picks up effects from all intermediate scales, arising as a cumulative effect, and relies 
on the ``depth'' of structure, i.e., on the ratio between the homogeneity scale and the scale of the smallest ripples
(recall from subsection~\ref{pathdependence} that no lower cut--off exists for the inhomogeneities in the GW formalism). Korzy\'nski's example thus illustrates the non--local nature of backreaction. In line with this he also has shown in Ref. \cite{korzynskiBH} that backreaction in a system of many compact sources can be large even if the metric is close to FLRW almost everywhere. The important issue there is the clustering properties of matter.

\section{Objections to the scalar averaging approach}
\label{sec:scalaraveraging}

The scalar averaging approach \cite{buchert:dust,buchert:fluid,buchert:jgrg} referred to here is a realisation of the thoughts advanced in Refs.\ \cite{ellis:average,ellisstoeger} (for other approaches, see Refs.\ \cite{brownetal,paranjapesingh,coley,maurokamilla,klingon}).  
In this section we address some misinterpretations of the scalar averaging framework that appear in Section 3 of \cite{gw}.

There are three points that must be considered: (i) the difference between spacetime notions versus a cosmological $3+1$ approach, (ii) the difference between the scalar averaging framework and metrical approaches to backreaction, and (iii) the importance of coordinate--independent arguments. 

Green and Wald also comment (in footnote 6 of \cite{gw}) that this approach has
the difficulty that the system of averaged equations is not closed. This (which we believe is the only relevant issue in Section 3 of \cite{gw}) has to be emphasized: even if one included a further evolution equation for the backreaction variable, as GW suggest, the system of averaged equations would not be closed, unless a dynamical equation of state is assumed or derived \cite{buchert:EOS}\footnote[11]{A second--order differential equation for the backreaction variable ${\cal Q}_{\cal D}$ (as the relativistic counterpart to Equation~\ref{q}) can be derived (T Buchert, unpublished), and it depends on further variables such as the intrinsic curvature backreaction appearing in metrical rescaling \cite{klingon}, implying the need to derive further equations for these new variables.  This situation is akin to the (infinite) moment hierarchies as they appear, e.g., in kinetic theory. A recent paper addressing closure illustrates this issue \cite{kasparsvitek}.}.   This issue has been spelled out at various places, e.g., it was already noted in the original paper \cite{buchert:dust}.  A closure of the system would be equivalent to saying that by shrinking the domain of averaging to arbitrarily small domains the equations would provide a general local solution of Einstein's equations. Rather, these are  conditions on averages similar to, e.g., the tensor virial theorem of Chandrasekhar and Lee \cite{chandralee} that also provides a balance relation between averaged variables by including fluctuations\footnote[12]{The tensor virial theorem is closed by the virial conjecture of stationarity of the averaged inertial tensor for an isolated system (a stationarity hypothesis can also be used in the scalar averaging framework, see Ref.\ \cite{buchert:static}). By construction, such balance relations cannot provide the local solutions.}.

We start with a common misunderstanding that also appears in \cite{gw}.

\vspace{-5pt}

\subsection{Can there be average acceleration with local deceleration everywhere?}

{\em Yes}, because backreaction is non--local.

Green and Wald state, and this opinion is shared by others since it {\em seems plausible}, that:
``... One can give an example of this sort \cite{ishibashiwald} wherein the `backreaction' is so large that {\em one obtains acceleration of the representative FLRW universe, even though each of the [disjoint]$\,$\footnote[13]{Literally, GW state ``disconnected'' and refer to Section 3 of \cite{ishibashiwald}, which concerns ``a model where at time $t$ the Universe consists of two disconnected(!) dust filled FLRW models\ldots''. Taken literally, topologically disconnected FLRW universes are physically less relevant.
Here, we discuss the more relevant case of {\em disjoint} domains whose union constitutes the whole spatial section. This remark applies to further citations from GW as in subsection~\ref{subsection:confusion}.} components of the actual Universe is decelerating}". (Our emphasis). 

The physical effect under question has often been explained in the literature (see, e.g., Refs. \cite{Rasanen:2006b}, \cite{buchert:focus}, subsection 5.2): the fraction of the volume of faster expanding regions grows more rapidly, so the average expansion rate rises.
Formally, we see this by looking at the Raychaudhuri equation
that governs the local expansion rate of dust matter: only a positive cosmological constant and vorticity can lead to an acceleration of the local expansion rate $\theta$ (in the sense that the time--derivative of $\theta$ is positive). Putting the cosmological term and the vorticity (which is active on small scales only) to zero, at first sight it  {\em seems implausible} that a collection of such decelerating fluid elements 
can lead to acceleration of some patch of the Universe. Let us look at the difference between the local and the volume--averaged Raychaudhuri equation for irrotational dust and vanishing cosmological constant:
\begin{equation}
\label{raychaudhuri}
\fl
\quad\dot{\theta} = - 4\pi G \varrho + 2{\rm II} - {\rm I}^2 \;\;\;\;;\;\;\;\;  \langle\theta\rangle^{\raise3pt\hbox{$\bolddot$}}_{\cal D} = - 4\pi G \average{\varrho} + 2\average{{\rm II}} - 
\average{\rm I}^2\;\;,
\end{equation}
where we defined the principal scalar invariants of the expansion tensor $\Theta_{ij}$, $2{\rm II}:= 2/3 \theta^2 - 2\sigma^2$ and ${\rm I}:=\theta$; averaging is performed on a spatial domain $\CD$, and we have used the commutation rule:
$\langle\theta\rangle^{\raise3pt\hbox{$\bolddot$}}_{\cal D} - \langle \dot{\theta} \rangle_{\cal D}= \langle \theta^2 \rangle_{\cal D} - 
\langle\theta\rangle^2_{\cal D} = \langle (\theta - \langle\theta\rangle_{\cal D})^2 \rangle_{\cal D}$.
Clearly, by shrinking the averaging domain to a point, the two equations agree. However, evaluating the local and averaged invariants,
\begin{eqnarray}
\fl
\quad
& 2{\rm II}- {\rm I}^2 = -\frac{1}{3} \theta^2 -2\sigma^2\;\;\;\;;\nonumber\\
\fl
\quad
& 2\average{\rm II} - \average{\rm I}^2 = - \frac{1}{3} \average{\theta}^2 - 2\average{\sigma}^2
+ \frac{2}{3} \average{(\theta - \average{\theta})^2} - 2 \average{(\sigma - \average{\sigma})^2} \;\;,
\end{eqnarray}
gives rise to two additional, positive--definite fluctuation terms, where that for the averaged expansion variance enters with a positive sign.

Thus, the time--derivative of an averaged expansion may be positive even if the time--derivative of the expansion is negative {\em at every point} in $\CD$. 
This is, technically, a consequence of the non--commutativity of averaging and time--evolution and, physically, of the non--local nature of averaging that takes correlations into account. 
Applying this fact to a model that is the union of disjoint FLRW submanifolds with spatial boundaries implies that the difference in expansion rates of the respective sections is a positive--definite expansion variance term (see references to such models in subsection~\ref{subsection:confusion}).

\vspace{-3pt}

\subsection{Can energy conditions be violated for the average dynamics?}
\label{subsect:energycondition}

\vspace{-2pt}

Again, the answer is {\em yes}.  It is a possible consequence of what has been said above. Green and Wald state a theorem that the 
{\em effective} energy momentum tensor $t_{ab}(0)$ has to obey the weak energy condition (see, e.g., Ref. \cite{gw3}, Eq. (2.8) in Theorem 2), where it is written in the abstract that ``the leading effect of small scale inhomogeneities on large scale dynamics is to produce a trace--less effective stress--energy tensor that itself satisfies the weak energy condition''. 
Green and Wald present in Ref. \cite{gw3} (section 4) an example for backreaction violating the weak energy condition (where they imply that this must be a consequence of  the violation of the local weak energy condition), and where ``the limiting metric has an effective stress--energy tensor which is not trace--less''. We comment on this example in \ref{appendixC}.

Green and Wald ``emphasize the importance of imposing energy conditions on matter in studies of backreaction'', but we emphasize that 
the {\em effective} stress--energy tensor may,
in general, violate energy conditions: the non--local fluctuation terms discussed above may lead to violation of energy conditions {\em upon averaging}, although {\em locally} they are satisfied. 

This has also been stressed in the literature with regard to the mapping of inhomogeneity effects to an effective scalar field that plays the role of a quintessence field in standard models for dark energy (see, e.g., Refs. \cite{buchert:review} and \cite{buchert:morphon}). While it is known that a {\em fundamental} scalar field (e.g., a quintessence field) must violate the weak energy condition and other physical properties (e.g., Ref. \cite{barbozanunes}) to explain observational data, an effective scalar field need not.
Since, as we have shown in subsection~\ref{noaverage}, the framework in which the problem is addressed by GW is {\em local}, the resulting effective stress--energy tensor would then have to obey the local energy conditions only, and only in the limit $\lambda \searrow 0$ (recall that there is no effective stress--energy tensor in the GW scheme for 
$\lambda \in (1,0)$), i.e., {\em their Theorem 2 does not apply to an averaged stress--energy tensor}. Addressing the real averaging problem would not deliver a local expression. We remind the reader that there is currently no agreed way to
average tensors in GR.

Green and Wald's theorem states that the effective stress--energy tensor emerging from 
a weak--limit of spacetimes obeying the Einstein equations has to obey the
weak energy condition. Such a result would be fine but irrelevant for backreaction.

\vspace{-3pt}

\subsection{Confusion between the scalar averaging framework and metric approaches}
\label{subsection:confusion}

\vspace{-2pt}

Green and Wald question the framework where backreaction is discussed
in terms of spatial averages  of scalar quantities. They assume that
such a procedure assigns an averaged FLRW metric
to the averaging region, and the ``main flaw with such approaches" is that the
metric thus obtained may be far from the real metric, even when the latter
is close to a FLRW metric, generating ``entirely spurious" backreaction terms \cite{gw}.

However, this criticism is based on a fundamental misinterpretation of the
scalar averaging formalism \cite{buchert:dust,buchert:fluid}:
it does not involve any notion of average metric (and does not refer to geodesic deviations \cite{gwnote}); only averaged scalars are considered. 

Explicitly, Green and Wald criticise the scalar averaging approach by considering two disjoint\footnote[14]{As stated above, we consider this property rather than the
original term ``disconnected''. We also interpret ``dust FLRW universe'' to refer to 
a spatially bounded submanifold of a dust FLRW universe.} dust FLRW universes in different stages of
expansion. In Ref. \cite{gw} they state ``The Buchert prescription would represent this [disjoint] Universe as a single
FLRW Universe, which provides a bad approximation to the actual metric everywhere.'' 
And: ``...the Buchert procedure with the above choice of hypersurface instructs us to represent the
Minkowski metric with an `averaged' FLRW metric $g (0)$ that is an extremely poor 
approximation to $g_{ab}$ on all scales." 
Again, we emphasize that in the scalar averaging framework there is no such ``instruction'' on a metric, apart from the spacetime split of the 4-metric (on which we comment below). As explained at the beginning of this section, the formalism defines spatial average properties among scalar variables that depend on second derivatives of the metric. The metric itself is not specified.

We note that there are indeed investigations in the literature that study so--called {\em template metrics} \cite{paranjapesingh}, \cite{morphon:obs}, \cite{rosenthalflanagan}, \cite{virialfraction}, \cite{testing}, that are intended to be compatible with the exact average properties (although these are not the papers referred to by GW). However, even these papers do not ``represent this [disjoint] Universe as a single FLRW Universe'' (the metric which the standard model exactly assumes). On the contrary, backreaction effects are often illustrated on the basis of models with two disjoint universe sections and the main effect is identified as being the result of the differences in expansion rates (see, e.g. \cite{wiltshire:exactsolution}, \cite{rasanen:peakmodel}, \cite{wiegandbuchert}, \cite{virialfraction}).

The FLRW metric itself can be viewed as a global template metric. If we find that the global average spatial curvature today is not zero, then the flat FLRW template will constitute a poor approximation of the spatial sections, and small deviations thereof will represent large deviations with respect to the physical background \cite{kolb:backgrounds}. This is exactly the problem that a template metric, even a single global template, is supposed to correct for. Moreover, as we argued in the context of the steel ball model analogy and the weak--limit framework: curvature inhomogeneities are not required to average out on an {\em assumed} background, as GW {\em a priori} impose to be true.

Green and Wald also argue that the average expansion rate $\average{\theta}$ is not physically meaningful, and that
deviations of $\average{\theta}$ from the FLRW value due to large ${\cal Q}_{\cal D}$ are ``entirely spurious'',
and that ``In realistic cosmological situations [...] the comoving synchronous hypersurfaces of the Buchert construction will provide a poor choice for approximating the hypersurfaces with nearly FLRW symmetry''. (Compare here our discussion of the backreaction term ${\cal Q}_{\cal D}$ as it is evaluated in Newtonian cosmology in subsection~\ref{subsect:newtonian}.)

It is well--known that the average expansion rate depends on
the choice of hypersurface, and the issue has been discussed at length
in the literature, where it has been argued
that the physically relevant averaging hypersurface
is the one of statistical homogeneity and isotropy, see e.g. 
\cite{Geshnizjani,Rasanen:2004,Kolb:2004,Kolb:2005da,Rasanen:2006b,Rasanen:2010a,rasanen:focus,buchert:focus,brownetal, Rasanen:2009b, Rasanen:2008b}.
It is irrelevant that there are hypersurfaces that are not
physically interesting, it only matters that averages on some hypersurfaces
give physically meaningful results and can be formulated in a covariant way.
(For discussion of covariance and gauge--invariance of scalar averaging, see Refs. \cite{veneziano1,veneziano2,smirnov}.)
The average expansion rate evaluated on some hypersurface is a
useful quantity so far as it gives an approximate description of what is observed, which is indeed the case for the hypersurface of statistical homogeneity and isotropy.
However, for realistic situations, the difference of averages taken on hypersurfaces of statistical homogeneity and isotropy 
from averages taken on the hypersurface generated by the fluid's rest frame
and measured by observers comoving with the matter component
of the Universe, well modelled by dust, is expected to be
negligible \cite{Rasanen:2009b}, but this should still be demonstrated in more detail.
A nontrivial example is provided by the Swiss Cheese model of
\cite{lavinto}, where the average expansion rate describes
the redshift (and to a lesser extent the luminosity distance) well,
even in a solution that is far from the FLRW case.
We emphasize that, in practice, observations of quantities such as the expansion rate
and density always involve spatial averages (or averages over null geodesics,
which can be related to spatial averages 
\cite{Rasanen:2008b, Rasanen:2009b,Bull:2012,lavinto}). {\em Averaging
is not a mathematical artifact, it is a feature of real observations
that has to be properly modelled}.

\vspace{-5pt}
\subsection{Coordinate dependence of Green and Wald's arguments: the comparison of metrics}
\label{subsection:coorddep}

The averaged metric may be a poor approximation to the $g (0)$ of the GW formalism (recall Eq.\ \ref{one}).  Even though the averaging formalism does not refer to an explicit metric, Green and Wald argue that this ``poor approximation'' is an issue. But it is misleading to base an argument on the fact that the metric {\em coefficient functions} are ``far from'' the values of a background metric.  It may be demonstrated that such an argument is coordinate dependent.  Consider as an example a {\em flat metric} and write it in different coordinates by introducing a diffeomorphism $x^a = f^a (X^b)$ such that the metric coefficients can be transformed to the Minkowski coefficients: $g_{ab} dX^a \otimes dX^b = \delta^M_{cd} f^c_{\; |a} f^d_{\; |b}dX^a \otimes dX^b = \delta^M_{cd} dx^c \otimes dx^d$ (the last equality uses the inverse of the transformation $f^a$, and a vertical slash denoting a partial derivative with respect to the local coordinates $X^a$).  The values of the coefficients $g_{ab} = \delta^M_{cd} f^c_{\; |a} f^d_{\;|b}$ may be ``far from'' the values of the Minkowski metric $\delta^M_{ab}$, but in fact they are the same metric arising from reparametrising a flat space.

We conclude that any comparison of metrics by just using metric coefficients is unphysical, since it depends on the coordinates used.
The question of ``how far'' one metric is from another is a geometrical question that might perhaps be addressed quantitatively through curvature invariants. Comparing two metrics in GR is quite an involved task: we can employ Cartan scalars to test for isometry of metrics, see, e.g., \cite{stephani,coleyetal}, but, as Cartan showed, this may require up  $n(n+1)/2$ covariant differentiations of the Riemann tensor in $n$ dimensions. (In four dimensional spacetimes, one only needs at most seven derivatives \cite{karlhede}.)
 
Such a test is justified by Theorem 9.1 of \cite{stephani} which tells us that the geometry of a sufficiently
smooth manifold is locally uniquely determined by the curvature {\em and
its derivatives} to some finite order. But (a) the arguments by GW do not include comparing derivatives and (b) we are not aware of any ``almost" version of
this theorem, i.e. a statement that, if in some region the curvature of
the metric $g$ and its derivatives are close, in a suitable topological
space, e.g., some Sobolev space ({\em cf.} \ref{appendixB}), to those of some metric $g(0)$, then $g$ is
close to $g(0)$ (again, in some suitable topological space). Note in this context (compare subsections~\ref{subsection:geroch} and \ref{pathdependence})
that GW's (and Burnett's) assumption $(ii)$, {\em cf.} subsection~\ref{subsection:scheme}, is in some coordinate components, so to be useful here the missing theorem would have to include something about how the
coordinates are chosen to get a limit as GW want and the GW prescription would have to say more about the derivatives of the curvature.

Another way to deal with the issue of closeness of metrics for spatial
averaging might be an appropriate generalization of the methods used
in \cite{Val05} which allow identification of the Schwarzschild metric
given the data on any suitably smooth hypersurface. 

Finally, we turn to the issue of whether there is any reason to think that the standard cosmological model, which is based on work nearly a century old \cite{einstein1917,lemaitre1927} (and unlike the standard particle physics model is the simplest conceivable cosmological solution), might not be the best description of our Universe.

\vspace{-3pt}
\section{Models and observations}
\label{sec:obs}

\vspace{-5pt}
An implicit assumption of the FLRW model is that the dynamics and observations of the inhomogeneous Universe 
can be modelled by the dynamics and evolution of a spatially homogeneous and isotropic universe model.
This procedure seemed to be completely adequate for many decades; indeed, unanticipated discrepancies between observations and the FLRW model such as the 
need for an accelerated homogeneous universe model \cite{Riess:1998cb,Perlmutter:1998np}, compare here Ref. \cite{bolejkoandersson}, are ascribed to a cosmological constant, or more generally dark energy with a possible $z$--dependent equation of state.  It is often stated that we are in an era of ``precision cosmology'' and that all observations are consistent with a ``standard'' cosmological model referred to as $\Lambda$CDM, which is an FLRW model with a stress--energy content of several components (cold dark matter, baryons, neutrinos, radiation, and a cosmological constant $\Lambda$) fitted to agree with observations.  

Another popular approach accounting for the apparent acceleration of the Universe is to assume the Einstein equations are not a good description of the Universe on cosmological scales and that ``modified gravity'' models are required to account for the observations. 

In this paper we remind the reader of another possibility.  We assume that the Einstein equations hold locally, but because of the inhomogeneous distributions of matter and geometry the FLRW model fails to adequately describe the observations.  While the FLRW--$\Lambda$CDM model may indeed be an adequate fitting model for many observations, 
future cosmological observations that are more accurate may provide evidence that
the FLRW--$\Lambda$CDM model is inadequate.
It may be that some of the many ``tensions'' between observations and the FLRW--$\Lambda$CDM model already constitute such evidence: there are several different observational challenges to the FLRW--$\Lambda$CDM
model at the 3$\sigma$ level \cite{Copi07,Copi13,Granett08ISW4sig,Flender13ISW3sig,HotNad2014JubISW,WiegBO14,
Battye14fivesigmaantiLCDM,Cyburt15Li7,Meneghetti13}, and some
observational questions have been raised about the identification of
the comoving rest frame on 60--100{\hMpc} scales---well above those
of strong--field astrophysical objects \cite{wsmw,RBOF15}.  These
results contrast with GW's statement that the FLRW--$\Lambda$CDM model is
in ``excellent agreement with all cosmological observations''.
However, this paper is
not intended as a discussion of potential observational problems
with (nor criticisms of) the standard model, but instead it analyses Green and Wald's theoretical claims.

We turn now to observational issues in dealing with the inhomogeneous Universe.

\subsection{Observational issues}

It is important to take into account large metric derivatives and corresponding changes of observational data interpretation. Green and Wald discuss this, underpinned by the claim that Newtonian notions fully capture these changes with the help of their dictionary, which is restricted to an assumed near--FLRW situation. 
We agree with their statement
``Moreover, the time evolution of the shear and convergence of a bundle of geodesics depends on the Riemann curvature (i.e., second
derivatives of the metric), which can be very large", but it is not clear that a ``Newtonianly perturbed FLRW model'' can fully account for this. This should be regarded as an open question until backed by concrete general calculations.
The issue is whether the actual GR solution or aspects of it can be constructed via a ``dictionary'' from a Newtonian solution (not shown), {\em not} whether a quasi--Newtonian GR solution can be constructed via some rules from a Newtonian solution. 

Green and Wald argue that a mapping of solutions of Newtonian
cosmological equations with periodic boundary conditions and certain properties
to GR solutions given in Ref. \cite{gw2} ``should yield an extremely accurate general relativistic description of our Universe, and, in particular, it provides the leading order backreaction effects at large scales produced by small scale inhomogeneities''.
However, there are two different issues. One is whether certain Newtonian
solutions are the limit of some GR solutions; 
the other is whether the GR solution that describes the real Universe is
at all times close to a corresponding Newtonian solution\footnote[15]{If this were
the case, backreaction would be small on the global scale, because in Newtonian theory its effect reduces to a boundary term
\cite{buchertehlers}, unlike in GR \cite{buchert:dust}.} (see also Ref. \cite{Todd2015}).

Ref. \cite{gw2} addresses the former question, but it
is the latter issue that is relevant for providing an accurate GR
description of our Universe and evaluating backreaction.
Even if a particular  GR solution starts from initial conditions close to 
a Newtonian solution, this does not imply that the GR solution would remain close to the Newtonian solution.
As a simple example, in Newtonian gravity an isolated two--body system
with an elliptic orbit is a stable configuration, whereas in GR the
orbit will decay, and the system will be driven far from the Newtonian
solution\footnote[16]{The global stability properties of a FLRW background have been investigated in--depth by a dynamical system analysis in the space of
physical backgrounds using scaling assumptions (i.e. backgrounds that emerge from the average of the inhomogeneous universe model) \cite{roy} (see also \cite{sussman:instability} for the class of LTB models). It has been found that the FLRW background is globally gravitationally unstable in two sectors, one corresponding to inhomogeneity effects mimicking dark energy behaviour on large scales, the other corresponding to inhomogeneity effects mimicking dark matter behaviour on small scales. In contrast to these instabilities the Newtonian inhomogeneities average out on the chosen background model by construction and this latter is stable \cite{buchertehlers}.}.
The difference in the evolution of the orbit is related to the
difference between the Weyl tensor and the corresponding
quantity in Newtonian theory, the Newtonian tidal tensor.
A general GR solution (even if the matter is dust) does not correspond to any
Newtonian solution. This is related to the fact that in Newtonian theory
the tidal tensor, corresponding to the electric part of the Weyl tensor,
does not have an evolution equation, and the tensor corresponding to
the magnetic part of the Weyl tensor is zero. In GR, both are in general
non--zero and have evolution equations, see Refs. \cite{Ellis:1971,Ehlers:1991,Ellis:1994,Kofman:1995,Matarrese:1995,rza1,EllisvanElst,vanElst:1998, Ehlers:1998, Szekeres,Ehlers:2009,Rasanen:2010a, Bertello:2012} for discussion.
In \cite{gw2} the arbitrariness of the tidal tensor is fixed by the assumption of periodic
boundary conditions. As the Newtonian equations are elliptic and do not
have a well--defined initial value problem, the boundary conditions
are essential \cite{buchertehlers}. This is quite different from GR, where there is
a well--defined initial value problem.
Changing the evolution of the boundary at distances much larger than
the GR particle horizon would not impact the GR solutions, but can
completely change the Newtonian solutions.
A lack of backreaction in GR cannot be established by starting
from the assumption that the Universe is well--described by Newtonian theory.

A common argument against backreaction, sometimes ascribed to \cite{weinberg} and repeated by GW is that 
``by flux conservation, the average of the apparent luminosity
of sources (including multiple images) must match the FLRW
value to a high degree of accuracy''.
However, if the area element is
different from FLRW, then the luminosity distance will also be different.
Green and Wald assert that it ``is a simple fact'' \cite{gwnote} that if two metrics
are perturbatively close, then the average apparent luminosity of sources
is a close match. Actually, this is incorrect. A counterexample is provided
by Enqvist et al. \cite{enqvist} in which, while 
perturbations (and their first derivatives) around an Einstein--de Sitter
background remain small, the luminosity distance can even match that of a $\Lambda$CDM
model with $\Omega_{\Lambda}=0.7$. It is also straightforward to construct exact spherically symmetric
counterexamples to the flux claim, such as the one in Ref. \cite{mustapha} (see also Refs. \cite{Ellis:1998a, Ellis:1998b}). 
This issue has been discussed in 
Section 2.1.2 of Ref. \cite{rasanen:peakmodel} and Section 4 of Ref. \cite{lavinto}.
Even on large scales statistical isotropy and homogeneity is not enough
to reduce the luminosity distance to its FLRW value, as shown in Ref. \cite{lavinto}. Indeed, the violation of the FLRW relationship
between the expansion rate and the luminosity distance can be used as a test of the importance of inhomogeneities
\cite{Clarkson:2007b, Rasanen:2008b, Shafieloo:2009, Rasanen:2009b, Boehm:2013,lavinto}.

\subsection{The (quint)essence of the backreaction approach}

As Green and Wald stress, second derivatives of the metric (i.e. spacetime curvature) have to be large in realistic models of the Universe. Averaging or smoothing these derivatives leads to an effective stress--energy tensor that is in general not trace--free, as we have demonstrated. Hence, this furnishes an argument {\em for} the potential importance of backreaction.

The Einstein equations dictate inhomogeneous curvature for inhomogeneous sources. A physical cosmology has to capture {\em both} inhomogeneities in the sources and inhomogeneities in the geometry. Idealising the latter by assuming a homogeneous FLRW metric globally, leads to a missing geometrical piece on the left--hand side of the Einstein equations that shows up as missing sources on the right--hand side of the Einstein equations in the standard model.
``Backreaction'' takes the inhomogeneities in geometry into account and so provides a more realistic average description of the left--hand side of the Einstein equations.

From first principles, any realistic cosmological model emerging from a $3+1$ spacetime split should evolve spatial curvature for evolving sources, even if the spatial curvature is set to zero at some initial time. The oversimplified standard model 
keeps curvature at this non--generic value, while backreaction models evolve curvature, even if it is set to zero at some initial time \cite{buchert:dust}. 

We have to distinguish average properties and fluctuations. 
Fluctuations may remain small on large scales, but that is not the issue: the issue is the background about which these fluctuations are small, and this is what is addressed by backreaction models.  Metrical deviations from an (unknown) average metric may be small, but it is an {\em assumption} that this average metric is dynamically equivalent to a homogeneous (and flat) FLRW solution. In other words, small metric perturbations do not imply small distortions of a flat geometry. 
In 1916, Einstein already clarified this and explained that the perturbations could equally well be small deviations from a large--scale curved space. (Einstein, however, reached this conclusion with calculations on which we comment, below his quote, from a modern perspective.) This can be considered a variant of the steel ball model of GW. In Einstein's words \cite{einstein}:
\begin{quotation}
{\em But it is conceivable that our Universe differs only slightly from a Euclidean one, and this notion seems all the more probable, since calculations show that the metrics of surrounding space is influenced only to an exceedingly small extent by masses even of the magnitude of our sun. We might imagine that, as regards geometry, our Universe behaves analogously to a surface which is irregularly curved in its individual parts, but which nowhere departs appreciably from a plane: something like the rippled surface of a lake. Such a Universe might fittingly be called a quasi-Euclidean Universe. As regards its space it would be infinite.
But calculation shows that in a quasi--Euclidean Universe the average density of matter would necessarily be nil. Thus such a Universe could not be inhabited by matter everywhere; it would present to us that unsatisfactory picture... If we are to have in the Universe an average density of matter which differs from zero, however small may be that difference, then the Universe cannot be quasi--Euclidean. On the contrary, the results of calculation indicate that, if matter be distributed uniformly, the Universe would necessarily be spherical (or elliptical). Since in reality the detailed distribution of matter is not uniform, the real Universe will deviate in individual parts from the spherical, i.e. the Universe will be quasi--spherical. But it will be necessarily finite}.
\end{quotation}
Einstein emphasized that the small deviations require a more careful description of the global spatial curvature. (Einstein mentions a positively curved
background, but it could as well be negatively curved.)
We argue that the deviations should be studied with respect to a background that is defined by the
actual average distribution (the {\em physical background}), as it is done in other physical disciplines that investigate fluctuation theories, e.g., in solid state physics,
not by {\em a priori} assumption.
 
In a modern perspective, but only in Newtonian cosmology, we can represent the Universe through a finite flat 3--torus model with Newtonian inhomogeneities on a factored out relativistic FLRW model, {\sl cf.} \cite{buchertehlers}.
In Newtonian cosmology the {\em a priori} assumption of a background that is equivalent with the average distribution can be realised: the spatially averaged model can be {\em assumed} to be an FLRW model on the imposed scale of the 3--torus due to the globally vanishing backreaction, which assures that the inhomogeneities average out to zero,
{\sl cf.} \cite{buchertehlers}; in general relativity such a construction is not possible, and the average model in general does not comply with an FLRW model \cite{buchertcarfora:curvature}:  the {\em physical background} will not coincide with an assumed fixed background, since it is 
determined through the dynamics of inhomogeneities.

Discussions restricting the regime of applicability to regions ``far from black holes or neutron stars'' overlook the fact that on cosmological scales the curvature can be tiny but still dominant over the source contributions.
Even in homogeneous cosmology, the Einstein cosmos \cite{einstein1917} provides an example of a curvature--dominated
model, its curvature radius being smaller than the Schwarzschild radius if the former is greater than about 3.3 Gpc (assuming a mean
density of the present estimate of the total baryonic plus dark matter densities and the observed value of the Hubble constant). 
In other words, this example shows that if the curvature is cosmologically significant small metric deviations from a flat metric would run into contradictions.

More generally, even if fields are weak then an important consideration
is the calibration of the asymptotic rulers and clocks of the close to
spatially flat metric around bound structures relative to a generally
non--FLRW cosmological average. Since no systems are truly isolated, this may
generally require a notion of ``finite infinity''
\cite{ellis:average,cox:finty,wiltshire:equiv,ellisstoeger:local}
to replace the exactly asymptotically flat geometry to which many definitions
of gravitational mass are tied. The problems of averaging and coarse--graining
may therefore be intimately related to fundamental unsolved problems
concerning gravitational mass--energy \cite{wiltshire:focus}.

Finally, in this paper we focussed on backreaction effects in general relativity, as this is also the framework addressed by Green and Wald in their papers.
However, there could be more degeneracies relating observations with theoretical predictions: we did not touch upon issues related to different theoretical frameworks such as the possibility of a non--symmetric connection, the role of torsion, non--metrical theories or, generally, modified gravitational theories that contain general relativity as a low--energy limit. A general assessment of backreaction effects would also incorporate studies of possible deviations from classical general relativity.

\vspace{-5pt}
\section{Conclusion}
\label{sec:conc}
\vspace{-5pt}

There is no proof that backreaction of inhomogeneities 
is irrelevant for the dynamics and observables of our Universe.  

\noindent
The most detailed study claiming to show that backreaction is irrelevant is in a series of papers by Green and Wald \cite{gw1,gw2,gw3,gw} and a precursor paper by Ishibashi and Wald \cite{ishibashiwald}, so we have examined those papers in detail (in particular the review in Ref.\ \cite{gw}).  We have pointed out several issues that lead us to conclude that these do not provide a proof that backreaction is irrelevant.  In particular, we have demonstrated that the claimed trace--free nature of backreaction is unphysical and is not relevant within the backreaction framework.   We have also shown that the GW framework does not address the actual question of backreaction since i) it is not concerned with averaging; ii) it does not include terms that should be present in a consistent treatment (especially, the trace parts); iii) its relation to the cosmological situation is unclear; iv) the example steel ball analogy is not concerned with the fitting problem; v) the ultra--local limit procedure does not take into account that backreaction is a non--local effect and vi) the criticisms misrepresent existing work on backreaction.

While the GW framework is not applicable to studying backreaction, it is possible that further developments of their framework may provide useful mathematical results.

\vspace{-5pt}
\section*{Remark on Green and Wald's note in response to this paper}
\vspace{-5pt}

Shortly after this paper was submitted, Green and Wald responded with a note \cite{gwnote} in which they
redefined the word ``backreaction'' to refer to the specific setting they address in their formalism, labelling other approaches---which
constitute most of the papers in the literature---as ``pseudo--backreaction''. They further clarified that their
formalism was ``never intended or claimed to apply'' to such approaches\footnote[17]{Since the previously published paper \cite{gw}
contains strong criticisms of such approaches without making this distinction clear, this paper constitutes a response to what has been
previously published, and this is not changed by any restatements of such issues in Ref. \cite{gwnote}. In response to Ref. \cite{gwnote} we have added
some clarifications and examples, particularly in \ref{appendixB}. These additions do not change any of our conclusions.},
and, in particular,  that their results ``do not apply to the Buchert formalism'' or to ``LTB models'' modeling large voids \cite{gwnote}.
Regardless of terminology, the physical question of interest is whether inhomogeneities have a significant effect on the evolution of averaged variables that correspond to observable quantities.
We have explicitly demonstrated that the definition of backreaction assumed by Green and Wald is too narrow to address this question.

\vspace{14pt}

{\em Acknowledgements:}
\small{The work of TB was conducted within the ``Lyon Institute of Origins'' under grant ANR--10--LABX--66. 
TB, JJO and BFR acknowledge support for a part of this project from the
HECOLS International Associated Laboratory, supported in part by the
National Science Centre, Poland, grant DEC--2013/08/M/ST9/00664, and
for a part of the project under grant 2014/13/B/ST9/00845 of the
National Science Centre, Poland. JJO acknowledges partial support from National Science Center, Poland, grant: Etiuda 2, the
Erwin Schr\"odinger International Institute for Mathematical Physics, Austria, and JJO and BFR acknowledge visiting 
support from CRALyon. AAC acknowledges support from NSERC. The work of EWK was supported by the Department of Energy through Grant No.\ DE--FG02--13ER41958.\\
Thanks to Krzysztof Bolejko,  Chris Clarkson, Ruth Durrer, Henk van Elst, Toshi Futamase, Miko{\l}aj Korzy\'nski, 
Sabino Matarrese, Slava Mukhanov, Todd Oliynyk, Roberto Sussman, Sebastian Szybka, Gabriele Veneziano and Herbert Wagner for their detailed reading of the
manuscript, constructive discussions and valuable remarks.  We also thank all of the referees for their helpful comments and acknowledge Bob Wald for discussions.}

\normalsize
\appendix

\section{On the intrinsic realisation of the steel ball analogy}
\label{appendixA}

In order to address the analogy given by GW, we have to investigate a well--defined intrinsic geometrical mapping between $(\mathbb{S}^2,\,g_{\mathrm pol})$ and $(\mathbb{S}^2,\,g_{\mathrm can})$. Even if we restrict ourselves to the two--dimensional case, as GW did---and we emphasized that this case is physically irrelevant for the purpose of investigating backreaction---we can exploit the Thurston--Rodin--Sullivan  $\epsilon$--approximation to the  Riemann mapping theorem  \cite{Andreev, He, Leibon, Rodin, Thurston, Thurston2} (associated with the circle packings techniques developed by W. Thurston). This is not a perturbation argument, but rather a  deformation argument in the space of (metric) geometries over $\mathbb{S}^2$. A more direct and easier variant of this would be the  Poincar\'e--Koebe  uniformisation theorem directly formulated for two--dimensional polyhedral metrics (the existence of this version of the standard uniformisation technique is a consequence of the fact that conical singularities are not seen by the conformal structure of  surfaces, and this has been nicely developed by many authors \cite{Troyanov, Troyanov2, Picard}). Both are mathematically well--defined and constructive procedures. They provide an explicit mapping (typically a conformal transformation) between $(\mathbb{S}^2,\,g_{\mathrm pol})$ and $(\mathbb{S}^2,\,g_{\mathrm can})$.  These mapping techniques are currently applied in the modelling of  two--dimensional discrete structures in fields ranging from image analysis to biology and medicine. In particular,  both approaches would map $(\mathbb{S}^2,\,g_{\mathrm pol})$ to a well--defined $(\mathbb{S}^2,\,g_{\mathrm can})$ without any conical defects whatsoever.

We would further like to correct the statement by Green and Wald that the metric of the polyhedron fails to be smooth also at the edges. The metric of $(\mathbb{S}^2,\,g_{\mathrm pol})$ is perfectly smooth at the edges. The bending we envisage at the edges, suggesting a lack of smoothness there, is an artifact of the embedding in Euclidean space $(\mathbb{E}^3,\,\delta)$. It is related to the (discretized) second fundamental form (the extrinsic curvature) of  $(\mathbb{S}^2,\,g_{\mathrm pol})$, and if we take it into account, then we need to consider the (discretized version) of the Gauss--Codazzi  constraint describing the compatibility conditions between the intrinsic and extrinsic geometry of  the  
embedding $(\mathbb{S}^2,\,g_{\mathrm pol})\hookrightarrow (\mathbb{E}^3,\,\delta)$, see, e.g., Refs. \cite{regge,carforamarzuoli}.

\vspace{-3pt}
\section{The nature of the weak--limit in the GW approach}
\label{appendixB}
\vspace{-3pt}

This is a technical appendix where we discuss the mathematical and geometrical underpinning of Green and Wald's approach. There are subtle assumptions and hypotheses underlying their claimed results that are unclear and not explicitly stated. Even when these hypotheses are spelled out clearly they are not always adequately implemented, eventually leading to insufficiently justified conclusions.  As a typical and important example, let us take the assumptions $(ii)$--$(iv)$, according to which \cite{gw1, gw} the components of the tensor fields $\gamma_{ab}(x,\lambda)$ and of $\nabla_c\gamma_{ab}(x, \lambda)$ are locally in $L_1$,  and bounded by smooth functions $\lambda\,C_1(x)$ and $C_2(x)$,  respectively,  see Eqs. (\ref{four}) and  (\ref{five}).  As often emphasized in \cite{gw, gw1}, no restrictions are placed upon second derivatives of $\gamma_{ab}$,  with the effect that \cite{gw} ``... second derivatives (i.e., the curvature) of $g_{ab}$ may have unbounded fluctuations relative to $g_{ab} (0)$.'' Nonetheless computations in \cite{gw} are performed as if the spacetime metrics considered are smooth, as is confirmed in the recent explanatory note \cite{gwnote}. Unfortunately, given the above hypotheses,
one cannot safely carry out computations by declaring that second derivatives of the perturbing tensor $\gamma$ may be unboundedly large, and then treat them as ordinary smooth functions.

In fact, $(ii)$--$(iv)$ imply that the GW effective stress--energy tensor $t_{ab}(0)$ cannot be defined in terms of the tensor 
field $\mu$ alone, (cf.~(\ref{six})): it may have components that in general make sense only as non--regular distributions. 
Furthermore, as we shall demonstrate below, under $(ii)$--$(iv)$ there is a very delicate regularity issue in taking a weak--limit, with the
consequence that one cannot prove that the trace of  $t_{ab}(0)$ vanishes without further hypotheses additional to those in $(ii)$--$(iv)$. In this sense, GW's Theorem 1 (the vanishing of the trace of the effective stress--energy tensor), is the counterpart of the observations we made for the steel ball analogy:
as in the analysis of the steel ball model, curvature fluctuations matter; their treatment requires an accurate analysis. 

In order not to have unstated assumptions and unclear formal computations, we address the mathematical analysis of GW's approach stepwise. To avoid the pitfalls of doing analysis with potentially unbounded quantities (here the second derivatives of the perturbation $\gamma$), we start by providing a Sobolev space analysis of the regularity issues associated with GW's assumptions $(ii)$--$(iv)$. We use Sobolev spaces to get to the point as quickly as possible and to pinpoint the origin of the problem lurking in the background.  Those not familiar with Sobolev space technology may skip this part of the analysis, if they are willing to accept the fact that taking weak--derivatives and weak--limits requires some caution when differentiating or integrating by parts.  In successive steps we address the technical limitations of GW's approach, which give rise to the stated difficulties in correctly defining their effective stress--energy tensor $t_{ab}(0)$. Many of these difficulties arise because 
of subtle issues in the adoption of the Burnett formalism, \cite{burnett} on which they heavily rely. 

\vspace{3pt}
\noindent
Step\; \emph{(i)}:\;To begin, we introduce, as do Green and Wald, a positive definite metric $e:=\,e_{ab}{\mathrm d}x^a\otimes {\mathrm d}x^b$ on the manifold $M$ with respect to which we can take local norms of tensors in defining the relevant functional spaces. In particular, for an open set $U\subset  M$ we denote by $L^p_{\mathrm loc}(U)$,\, $1\leq p \leq \infty $, the space of Borel measurable functions $f$ defined on $U$ which are $p$--summable, \emph{i.e.}, $||f||_{L^p(K)}\,:=\,\int_K\,|f(x)|^p\,{\mathrm d}\mu_{e}(x)\,<\,\infty $ for every compact set $K\subset  U$, where ${\mathrm d}\mu_e$ denotes the Riemannian measure associated with the metric $e$. For $p=\infty$, we denote by $L^\infty_{\mathrm loc}(U)$ the Banach space consisting of the essentially bounded functions on $U$, endowed with the essential supremum norm $||f||_{\infty }\,:=\,\inf\,\left\{C\in\mathbb{R}\,:\,f\,\leq\,C\;\mbox{almost everywhere in U}  \right\}$.
We also introduce the class of functions in $L^p_{\mathrm loc}(U)$ whose 
weak first derivatives, (see below for definitions), are also $L^p_{\mathrm loc}(U)$--functions. These are the Sobolev spaces $W_{\mathrm loc}^{1,p}(U)$\, $1\leq p \leq \infty $,\,often used in geometric analysis  and mathematical general relativity. Their definition naturally extends to the appropriate spaces of tensor fields. For further details see, e.g., Refs. \cite{Aubin, Badr, Choquet, Evans}.

It is also appropriate to recall the definitions of distributional and weak derivatives. Entering into such mathematical detail may appear pedantic. However, the different properties of these two derivatives play a subtle, but fundamental, role in what follows. 
As usual, let $C^\infty_0(M)$ be the space of smooth compactly supported functions on~$M$.  If we denote by $T_f$ the regular distribution associated with $f\in L^1_{\mathrm loc}(U)$, i.e., $T_f(\phi):=\,\int_M\,f(x)\,\phi(x)\,{\mathrm d}\mu_{e}(x)$, for $\phi\in C^\infty_0(M)$, then $T_f$ always admits a distributional derivative $\nabla\,T_f$, defined according to $\nabla\,T_f(\phi)\,:=\,-\,T_f(\nabla\,\phi)$, \emph{viz.}.
\begin{equation}
\nabla\,T_f(\phi)\,:=\,-\,\int_M\,f(x)\,\nabla\phi(x)\,{\mathrm d}\mu_{e}(x)\,,\,\,\,\forall \,\phi\,\in\,C^\infty_0(M)\;.
\end{equation}
If $f$ is continuously differentiable, we can integrate by parts on the right--hand side,
\begin{equation}
\fl\qquad
\nabla\,T_f(\phi)\,:=\,-\,\int_M\,f(x)\,\nabla\phi(x)\,{\mathrm d}\mu_{e}(x)
=\int_M\,\nabla f(x)\,\phi(x)\,{\mathrm d}\mu_{e}(x)=\,T_{\nabla f}(\phi)\;,
\end{equation} 
which provides the usual, often abused, rationale in manipulating distributional derivatives. Higher--order distributional derivatives are defined in a similar way. In nonlinear problems, like the one we are discussing here, it is more useful and sometimes indeed necessary to consider, rather than the distributional derivative, the more restrictive concept of weak derivative. Let $f\in L^1_{\mathrm loc}(U)$ be a locally integrable function on the open set $U\subseteq M$ and let $T_f$ be the corresponding regular distribution. If there exists a locally integrable function $F \in L^1_{\mathrm loc}(U)$ such that 
$\nabla\,T_f\,:=\,T_F$,\,\emph{viz.}
\begin{equation}
\label{defWD}
\fl\qquad
\int_M\,f(x)\,\nabla\phi(x)\,{\mathrm d}\mu_{e}(x)\,=\,-\,\int_M\,F(x)\,\phi(x)\,{\mathrm d}\mu_{e}(x)\,,\,\,\,\forall \,\phi\,\in\,C^\infty_0(M)\;,
\end{equation}
then we say that $F$ is the weak derivative of $f$, and write $F=\nabla\,f$. Weak derivatives emphasize local summability, whereas distributional derivatives put the accent on differentiability. The two notions agree for smooth functions, and often the notion of weak derivative is tacitly traded for the definition of derivative in the sense of distributions. But the two have quite different properties. In particular, the weak derivative may not exist: two well--known examples are the Heaviside step function and the Cantor function. Neither admits a weak derivative whereas they both do have a distributional derivative: the Dirac measure supported at the origin, and the Lebesgue-Stieltjes measure supported on the Cantor set, respectively. Both of these distributions are non--regular distributions, i.e., they cannot be associated with a locally summable function. Note also that the existence of a weak derivative is not equivalent to the existence of a pointwise derivative almost everywhere. 

Given these technical issues one may ask why we should not simply interpret GW's formalism in the distributional sense, as
suggested by their use of the weak topology (i.e., weak--limits and
integration by parts), and confirmed by their recent explanatory note \cite{gwnote}. To illustrate why not, we will present a simple example to show that there is a price to pay if one goes distributional without due care, in particular when interpreting weak--limits as a form of averaging. The example in question is related to the vanishing of the weak--limit $\wlim_{\lambda\rightarrow 0}\,\nabla_a \nabla_b \gamma_{ce}(\lambda)$, a basic result which has many implications in GW's approach.
 
For simplicity, let us work on $\mathbb{R}$, and consider the function (playing the role of GW's metric perturbation $\gamma_{ab}(x, \lambda)$)
\begin{equation}
\label{example1}
\gamma(x, \eta)\,:=\,\frac{\eta^2}{\sqrt{4\pi\eta}}\,e^{\,-\,\frac{x^2}{4\eta}}\,=\,\eta^2\,p_\eta(x, \eta;\,0)\;,
\end{equation}
where $p(x, \eta;\,0)$ is the heat kernel in $\mathbb{R}$ with source at $x=0$, and  where $\eta\in(0,1]$ (i.e., we are taking $\lambda:= \eta^2$).  One has
\begin{equation}
\fl\quad
\frac{\mathrm d}{{\mathrm d}x}\,\gamma(x, \eta)\,=\,-\,\frac{x}{2}\,\frac{\eta}{\sqrt{4\pi\eta}}\,e^{\,-\,\frac{x^2}{4\eta}}\quad;\quad
\frac{{\mathrm d}^2}{{\mathrm d}x^2}\,\gamma(x, \eta)\,=\,\left( \frac{x^2}{4}-\frac{\eta}{2}\right)\,\frac{1}{\sqrt{4\pi\eta}}\,e^{\,-\,\frac{x^2}{4\eta}}\;.
\end{equation}
Clearly, for all $\eta>0$, we have smooth functions, i.e., $\gamma(x, \eta)$,\,$\frac{\mathrm d}{{\mathrm d}x}\,\gamma(x, \eta)$, and $\frac{{\mathrm d}^2}{{\mathrm d}x^2}\,\gamma(x, \eta)$ are $C^{\infty}(\mathbb{R})$. Moreover, $0<\,\gamma(x, \eta)\,\leq\,\eta^2\,(4\pi\,\eta)^{-\,1/2}$, and $0<\,|\frac{\mathrm d}{{\mathrm d}x}\,\gamma(x, \eta)|\,\leq\,\eta\,(4\pi\,\eta)^{-\,1/2}\,\frac{|x|}{2}$, hence $\gamma(x, \eta)$ tends to zero as $\eta\searrow 0$, uniformly in $x$, while $\frac{\mathrm d}{{\mathrm d}x}\,\gamma(x, \eta)\rightarrow 0$ pointwise.  Finally,  since $\lim_{\eta\searrow 0}\,p_\eta(x, \eta;\,0)$ is not locally summable, $\frac{{\mathrm d}^2}{{\mathrm d}x^2}\,\gamma(x, \eta)$ tends to zero in the sense of distributions. Explicitly, let $T_{ ({x^2}/{4}-{\eta}/{2})}$ and $T_{p(x, \eta;\,0)}$ be the regular distributions associated, for any fixed $\eta\in (0, 1]$, with the smooth functions  $ (\frac{x^2}{4}-\frac{\eta}{2})$ and $p(x, \eta;\,0)$.  Since, for any $\phi \in C_0^{\infty} (\mathbb{R})$, $ (\frac{x^2}{4}-\frac{\eta}{2})\,\phi\,\in C_0^\infty(\mathbb{R})$,  and $\lim_{\eta\searrow 0}\,\int\,x^2\phi(x)\,p(x, \eta;\,0)\,{\mathrm d}x\,=\,(x^2\phi(x))|_{x=0}\,=\,0$ (by the known properties of the heat kernel $p(x, \eta;\,0)$), we can write  
\begin{equation}
\label{2derweak}
\lim_{\eta\searrow 0}\,T_{ ({x^2}/{4}-{\eta}/{2})}\,T_{p(x, \eta;\,0)}\,=\,\frac{x^2}{4}\,{\delta}_0\,=\,0\;,
\end{equation}
where $\delta_0$ is the Dirac measure supported at $x=0$. This shows that using
distributions is a far cry from what we would naturally consider, i.e., the vanishing of $({\mathrm d}^2/{\mathrm d}x^2)\gamma$ resulting from a suitable \emph{average} of a locally summable function over the region of interest\footnote[18]{In Green and Wald's papers it is often suggested that the weak--limit ``corresponds roughly to taking a local spacetime average''. In particular, the concluding paragraph of \cite{gw}  strongly emphasizes this aspect.}.  The above example clearly shows that, rather than an average, a direct distributional approach may give rise to measure concentration phenomena (associated here with the heat kernel measure $\frac{1}{\sqrt{4\pi\eta}}\,e^{\,-x^2 / 4\eta}\,{\mathrm d}x$  concentrating, as $\eta\searrow 0$, the zero weak--limit of $({\mathrm d}^2/{\mathrm d}x^2) \gamma$ at the origin $x=0$).

One may argue that, in contrast to the pathological
example (\ref{example1}), the \emph{oscillating to death} function $\gamma(x, \eta)\,=\,\eta\,\sin(x/\eta)$, $0\,<\,\eta\,\leq\, 1$,  often quoted in \cite{gw}, is a simple example modeling the assumptions $(ii)$--$(iv)$, with a nicer behaviour.  In this case we have  $\frac{d\gamma(x,\eta)}{dx}=\cos(x/\eta)$,  and   $\frac{d^2\gamma(x,\eta)}{dx^2}=-\frac{1}{\eta}\sin(x/\eta)$.   Since  $\int_{-a}^{a}\,\sin(x/\eta)\,{\mathrm d}x$ is uniformly bounded, one immediately gets $\lim_{\eta\searrow 0}\,T_{{{\mathrm d}^2\gamma(x,\eta)}/{{\mathrm d}x^2}}=0$ as a regular distribution. Moreover,  one easily computes 
\begin{equation}
\label{noncont}
\lim_{\eta\searrow 0}\,T_{\gamma(x,\eta)}\,T_{{d^2\gamma(x,\eta)}/{dx^2}}\,=\,-\,\frac{1}{2}\;,
\end{equation}
which, while indeed providing an elementary realization of GW's assumption $(iv)$, also shows that modeling $\gamma_{ab}(x,\lambda)$ and its derivatives (hence $g_{ab}(x,\lambda)$) as regular distributions---complying 
with $(ii)$--$(iv)$---is quite problematic: there can be a lack of continuity in the relevant geometric operators\footnote[19]{The relation (\ref{noncont}) is a standard example proving that multiplication is not a continuous operation in distribution space. When computing curvatures for $g_{ab}(x,\lambda)$ we need to use the product $T_{\gamma(x,\eta)}\,T_{{{\mathrm d}^2\gamma(x,\eta)}/{{\mathrm d}x^2}}$.  }. In particular, the mapping that associates to the metric $g_{ab}(x,\lambda)$ its Riemannian curvatures, say  the Riemann tensor $g_{ab}(x,\lambda)\longmapsto {\mathcal{R}^d}_{abc}(x,\lambda)$, cannot be a continuous map between the appropriate space of tensor fields, if the weak--limit and derivatives are interpreted distributionally without due care. 
For these reasons it is more exact to use 
weak rather than distributional derivatives. Even in this case distributions may nonetheless arise, as we shall see in the computation of the Einstein tensor for low regularity metric perturbations such as those occurring in GW's case. However, our calculation will be done in a controlled way from the point of view of geometrical analysis, allowing us to pinpoint the origin of the potential problems in the formalism.

Green and Wald's assumptions $(ii)$--$(iv)$ imply that each metric tensor $g(\lambda)$ along the curve of metrics (\ref{two}), is a low regularity (perturbation of a) Lorentzian metric.  The minimum assumptions needed to define the Riemann tensor (and all attendant curvatures) along (\ref{two}),  and at the some time comply with GW's assumptions $(ii)$--$(iv)$, are to assume that on the smooth background provided by $(M,\,g (0))$ the components of the tensor field $\gamma_{ab}(\lambda)$ are  essentially bounded and, together with their derivative, are locally square summable. Indeed, these are the minimal technical requirements needed to safely carry out most of the formal computations in Green and Wald's papers.  Explicitly, we require that around the generic point in $(M,\,g (0))$ there exists a  local coordinate neighbourhood $(U,\,x^a)$, $U\subset M$, where we can write
\begin{equation}
g_{ab}(\lambda)\,=\,g_{ab}(0)\,+\,\gamma_{ab}(\lambda)\;,\;\;\;\;\;\lambda\in(0,1]\;,
\end{equation}
and where the components of the perturbation tensor field $\gamma$  are such that
\begin{equation}
\gamma_{ab}(\lambda,x)\,\in\,W_{\mathrm loc}^{1,2}(U)\,\cap \,L_{\mathrm loc}^\infty(U)\;,
\end{equation}
for all $\lambda\,\in\,(0, 1]$. Here, according to the notation introduced above, $W_{\mathrm loc}^{1,2}(U)$ is the (local) Sobolev space of sections which together with their first derivative are square summable (in $U$), and  $L_{\mathrm loc}^\infty(U)$ denotes sections which are essentially summable in $U$, 
(the relevant Sobolev norms on $(M,\,g (0))$ being defined in $(U,\,x^a)$ \cite{Aubin, Badr, Choquet}). Note that GW also impose the \emph{smallness} of  $\gamma_{ab}(\lambda,x)$ as $\lambda\searrow 0$, by assuming that $|\gamma_{ab}(\lambda)|\,\leq\,\lambda\,C_1(x)$ for some smooth positive function $C_1(x)$ on $M$. In our setting this is just a smallness constraint on the local components of $\gamma$ and does not imply the smallness of the $W_{\mathrm loc}^{1,2}(U)$--norm of $\gamma$, in line with the GW formalism. 

It is convenient to take $(M,\,g (0))$ to be a generic background metric (similar to the case of the almost--everywhere flat metric in the steel ball analogy).  This is in line with the remarks in GW that their computations hold for generic background metrics. By so doing we can better appreciate the differences arising from assuming FLRW as the background. 

The Sobolev space assumption above provides a  local control on $\gamma$ and its first derivatives that is  consistent with the simpler GW hypotheses  $(ii)$--$(iv)$, but at the same time allows us a more precise analytical control on the geometrical tensor fields we need to use. To wit, since the set  $W_{\mathrm loc}^{1,2}(U)\,\cap \,L_{\mathrm loc}^\infty(U)$ is an algebra under pointwise multiplication \cite{Badr}, we can define all algebraic manipulations\footnote[20]{We can define all algebraic manipulations in the algebra associated with the product of the equivalence classes of
functions defined almost everywhere.} of $g(\lambda)$ and $\gamma(\lambda)$. Moreover, $\nabla \gamma(\lambda)$ and the Christoffel symbols $\Gamma^c_{ab}(\lambda)$ of   $g(\lambda)$  are in $L^2_{\mathrm loc}(U)$. More precisely, from the standard formula relating the Christoffel symbols of the two metrics $(M,\,g^{(0)})$ and $(M,\,g(\lambda))$ we have
\begin{equation}
\fl
\Gamma^a_{bc}(\lambda)\,=\,\Gamma^a_{bc}(g (0))\nonumber\\
-\,\frac{1}{2}\,\nabla_k g^{ij}(\lambda)\,
\left( g_{bi}(\lambda)g_{cj}(\lambda)g^{ak}(\lambda)-g_{bi}(\lambda)\delta_c^k\delta_j^a -
g_{ci}(\lambda)\delta_b^k\delta_j^a \right),\nonumber
\end{equation}
and we can symbolically write, in terms of  the  functional class,
\begin{equation}
\label{smoothpart}
\Gamma^a_{bc}(\lambda)\,\in\,C^{\infty}(U)\,+\,L^2_{\mathrm loc}(U)\;,
\end{equation}
where $C^{\infty}(U)$ here and in the following formulas refers to the smooth part of the relevant quantity under discussion, associated with the background $(M,\,g^{(0)})$\footnote[21]{For a similar treatment of low regularity metrics in Riemannian geometry, see \cite{Liimatainen}.}, (assumed to be smooth).

Since  $\nabla \gamma(\lambda)\,\in\,L^2_{loc}(U)$, it follows that $\nabla\gamma(\lambda)\,\nabla\gamma(\lambda)$ is in  
 $L^1_{\mathrm loc}(U)$, i.e., it is locally summable for all $\lambda\in (0,1]$ and hence there exists the weak--limit tensor $\mu_{abcdf}$ defined by (\ref{six}) as a regular distribution locally represented by a (tensor--valued) continuous function. In general, this is not a smooth tensor field, as assumed in \emph{(iv)}. However,  the smoothness requirement is not really relevant to the GW argument since the terms causing trouble are the second derivatives of $\gamma$. 
Indeed, from the (formal) local definition of the Riemann tensor components in $(U, x^a)$, 
\begin{equation}
R^d_{\;abc}(\lambda)\,:=\,\partial_a\Gamma^d_{bc}(\lambda)-\partial_b\Gamma^d_{ac}(\lambda)\,+\,
\Gamma^m_{bc}(\lambda)\Gamma^d_{am}(\lambda)\,-\,\Gamma^m_{ac}(\lambda)\Gamma^d_{bm}(\lambda)\,,\nonumber
\end{equation}
it follows that the terms $\partial_a\Gamma^d_{bc}-\partial_b\Gamma^d_{ac}$ in  
$R^d_{\;abc}(\lambda)$, containing the  second derivatives of $\gamma$, can be given sense only as elements of the distributional space $W_{\mathrm loc}^{\,-\,1,2}(U)$ (roughly speaking, the topological dual of  $W_{\mathrm loc}^{1,2}(U)$), whereas the quadratic terms containing  the first derivatives of  $\gamma$,\, \emph{i.e.}\;$\Gamma^m_{bc}\Gamma^d_{am}\,-\,\Gamma^m_{ac}\Gamma^d_{bm}$ are, as introduced above, in $L^1_{\mathrm loc}(U)$. Clearly, since $(M,\,g (0))$ is assumed smooth, in the second--order linear part of the Riemann tensor,   $\partial_a\Gamma^d_{bc}-\partial_b\Gamma^d_{ac}$,  there is a smooth part  (cf. \ref{smoothpart}), and  we can  write
\begin{equation}
R^d_{\;abc}(\lambda)\,\in\,W_{\mathrm loc}^{-1,2}(U)\,+\,L^1_{\mathrm loc}(U)\,+\,C^{\infty}(U)\;.
\end{equation}
Since contraction is an algebraic operation, the same situation holds also for the Ricci tensor, and  we have
\begin{equation}
R_{ac}(\lambda)\,:=\,R^d_{\;adc}(\lambda)\,\in\,W_{\mathrm loc}^{-1,2}(U)\,+\,L^1_{l\mathrm oc}(U)+\,C^{\infty}(U)\;.
\end{equation}
The situation for the scalar curvature is subtler: in order to contract the Ricci tensor we need to \emph{trace} the distribution--valued tensor components  $R_{ac}(\lambda)$ with the components $g^{ab}(\lambda)$ of the inverse metric, (which are in  $W_{\mathrm loc}^{1,2}(U)\,\cap \,L_{\mathrm loc}^\infty(U)$). For the components of the Einstein tensor we have
\begin{equation}
G_{ab}(\lambda)\,\in\,W_{\mathrm loc}^{\,-1, 2}(U)+\,C^{\infty}(U)\quad,\quad {\rm for}\;\; \lambda\,\in\, (0,1] \;\;.
\end{equation}

\vspace{7pt}
\noindent
Step\; \emph{(ii)}:\; Clearly, what makes the components $G_{ab}(\lambda)$ of the Einstein tensor potentially described by singular distributions is the presence of the $\nabla \nabla \gamma(\lambda)$ terms. Since we are dealing with a family of metrics $\lambda\longrightarrow g(\lambda)$, $\lambda\,\in\,(0,1]$, it may well happen that the distributional part of $G_{ab}(\lambda)$ vanishes in the limit $\lambda \searrow 0$ as a consequence of the GW hypotheses $(ii)$--$(iv)$. Even if this is the case, in order to give substance to GW's formalism, we must carefully check that this  vanishing weak--limit is supported on a set of significant
measure and not associated with a delta--like measure or with a discontinuous dependence on the averaging
scale (see the example (\ref{example1})--(\ref{noncont})).
In fact, Green and Wald claim that with the weak--limit operation many of these second derivative terms over the smooth background $(M,\,g (0))$  vanish in the limit $\lambda\searrow 0$. 
In particular, they assume in \cite{gw1} that the terms $g_{ik}(0)\,\nabla _a\nabla_b \gamma_{ef}(\lambda)$ weakly vanish in the $\lambda\searrow 0$ limit. Explicitly, by referring to the terms containing two derivatives of $h_{ab}(\lambda)$, they state\footnote[22]{Note: $h_{ab}(\lambda)$ of \cite{gw1} is $\gamma_{ab}(\lambda)$ here and in \cite{gw}.} just before formula (13) in\cite{gw1} ``These terms can be classified into the following types: (a) terms linear in $h_{ab}(\lambda)$, corresponding to the linearized Einstein operator acting on $h_{ab}(\lambda)$; (b) terms quadratic in $h_{ab}(\lambda)$, corresponding to the second--order Einstein operator acting on $h_{ab}(\lambda)$; and (c) terms cubic and higher--order in $h_{ab}(\lambda)$. The weak--limit of terms of type (a) vanish by the type of argument leading to (3)....". Actually, in claiming that the terms of type (a) vanish by the type of argument leading to (3), they implicitly extend Burnett's argument \cite{burnett} (as exploited in Eq. (3) of \cite{gw1} for first derivatives of $\gamma_{ab}$) to second derivatives, an extension that under their stated hypotheses is not guaranteed.

Explicitly, and regardless of the potential distributional nature of the Einstein tensor discussed above, the problem in GW's approach lies in analysing the weak--limit of their equation (12) in Ref. \cite{gw1}. This is the core of their work, but unfortunately, the analysis there is performed with some hidden assumptions. Given the importance of this point, let us describe it in detail.
What GW exploit in proving the vanishing of the weak--limit of terms of type (a) is the direct application of case (b) of ``Burnett's Theorem'' by identifying Burnett's tensor field $\alpha(\lambda)$ with $\nabla \gamma_{ab}(\lambda)$, ({\sl cf}. Ref. \cite{burnett}, pp.95--96, and in particular the top of p.92 where Burnett claims that ``$\nabla\,C^c_{ab}(\lambda)\,\longrightarrow \,0$ weakly").  Green and Wald implicitly assume that, for any compactly supported test tensor field density $f^{abce}$, one can write
\begin{equation}
\fl\qquad
\wlim_{\lambda\searrow 0} \nabla _a\nabla_b \gamma_{ce}(\lambda)
=\, \, \liml  \int f^{abce} \nabla_a \nabla_b \gamma_{ce} = -\liml \int \nabla_a f^{abce} \nabla_b \gamma_{ce}\, = \, 0\;,
\label{weakassumption}
\end{equation}
where the first step is  integration by parts of the term $\nabla _a\nabla_b \gamma_{ce}(\lambda)$,  and where in the second step one exploits the weak convergence to $0$ of $\nabla_b \gamma_{ce}(\lambda)$.

This would be fine, 
if we interpreted (\ref{weakassumption}) in the distributional sense and exploit the continuity of the distributional derivative. However, as shown above, this may generate a  vanishing weak--limit associated with concentration phenomena or exhibiting a discontinuous dependence on the averaging scale.
To avoid this we need to require that in the limit $\lambda \searrow 0$,  the tensor components $\nabla _a\nabla_b \gamma_{ce}(\lambda)$ are represented by functions which are at least locally summable. But this strict weak derivative representation is also problematic, since it is not {\em a priori} allowed under the  general assumptions on $\gamma_{ce}(\lambda)$ adopted by GW, in particular if $\nabla _a\nabla_b \gamma_{ce}(\lambda)$ can be unboundedly large. This indicates that local summability may also be a very delicate issue in GW's formalism and cannot be taken for granted. In particular, in order to define the quantities $\nabla _a\nabla_b \gamma_{ce}(\lambda)$ in the weak sense and give meaning to (\ref{weakassumption}) without nonregular distributions coming in, we have to assume, according to the very definition  (\ref{defWD})   of the weak derivative, that for every $\lambda\in (0,1]$ there is an array of functions $F_{abce}(\lambda)=(F_1(\lambda),\ldots,F_n(\lambda))$, each of which is in $L^1_{loc}$, such that
\begin{equation}
\int\,F_{abce}(\lambda)\,f^{abce}\,=\,-\,\int \nabla_b \gamma_{ce}(\lambda)\,\nabla _a\,f^{abce}\;,
\label{weak1}
\end{equation}
for all compactly supported test tensor densities $f^{abce}$. These $F_{abce}(\lambda)$ are the weak derivatives of $\nabla_b \gamma_{ce}(\lambda)$, namely we can symbolically write $F_{abce}(\lambda)\,:=\,``\nabla _a\nabla_b \gamma_{ce}(\lambda)"$.  In the GW framework there are  further constraints in addition to local summability that $F_{abce}(\lambda)$ must then comply with. Since 
\begin{equation}
\wlim_{\lambda\searrow 0}  \nabla_b \gamma_{ce}(\lambda)\,=\,0\;,
\end{equation}
the weak characterisation (\ref{weak1}) of  $\nabla _a\nabla_b \gamma_{ce}(\lambda)$ and the GW assumptions \emph{(ii), (iii), (iv)} necessarily require that the locally summable functions $F_{abce}(\lambda)$ must have a vanishing weak--limit as $\lambda\searrow 0$.
Moreover, we have to artificially require that this weak convergence to zero  of $\nabla _a\nabla_b \gamma_{ce}(\lambda)$ is not uniform in $\lambda$, i.e., we must require that $\nabla _a\nabla_b \gamma_{ce}(\lambda)$ are not uniformly bounded as $\lambda$ varies. This latter constraint is necessary in order to have that 
\begin{equation}
\wlim_{\lambda\searrow 0}\,\gamma_{cd}(\lambda)\,\nabla _a\nabla_b \gamma_{ef}(\lambda)\,=\,
-\mu_{abcdef}\;,
\label{tricky}
\end{equation}
({\sl cf}. Eq. (13) in Ref. \cite{gw1}). It is easy to see that, if  the weak convergence of $\nabla _a\nabla_b \gamma_{ce}(\lambda)$ to zero were uniform, then the weak--limit (\ref{tricky}) would be $0$. 

This non--uniformity, (more or less tacitly assumed in GW's \emph{(ii), (iii), (iv)}), is again an argument that GW \cite{gw1} take from Burnett ({\sl cf}. the displayed equation on top of page 92 of Ref. \cite{burnett}). In Burnett's words \emph{``Does $(g_{de}(\lambda)-g_{de}(0))\,\nabla_m\,C^c_{ab}(\lambda)\,\longrightarrow \,0$ weakly as $\lambda\longrightarrow 0$?\; Not in general!\, Although $g_{de}(\lambda)-g_{de}(0)\,\longrightarrow \,0$ uniformly and $\nabla_m\,C^c_{ab}(\lambda)\,\longrightarrow \,0$\, weakly, $\nabla_m\,C^c_{ab}(\lambda)$ need not be uniformly bounded. In fact, if $g_{ab}(\lambda)$ also satisfies condition (iv), then $(g_{de}(\lambda)-g_{de}(0))\,\nabla_m\,C^c_{ab}(\lambda)$ converges weakly to some expression in $\mu_{mnabcd}$.\,..."}. 

Whereas this lack of uniformity makes sense in the high--frequency limit in gravitational wave theory---simply because non--uniformity is almost intrinsic to wave propagation in the high--frequency regime (typically modelled after the behaviour of the function $\lambda\,\sin\,\left(x / \lambda \right)$)---it seems unlikely to be justified in cosmological backreaction where there are finite averaging scales defined by the gravitational system considered: what does ``high--frequency'' mean in this setting?  Is this ultra--local limit sensible in cosmology?

In a realistic cosmological averaging procedure, we may allow very high density contrasts $\delta\rho/\rho\,\gg\,1$, (hence very large $\nabla _a\nabla_b \gamma_{ce}(\lambda)$), but we have to control the averaging scale over which this oscillating contrast is relevant and hence assume a uniform boundedness hypothesis on $\nabla _a\nabla_b \gamma_{ce}(\lambda)$. Otherwise, if we want to have potentially unboundedly large $\nabla _a\nabla_b \gamma_{ce}(\lambda)$ while at the same time avoiding the  distribution--valued curvature that, as we have seen, naturally develops under such hypotheses, we do need, as in GW's formalism, a rather artificial fine--tuning between the non--uniformity of the upper bound of $|\nabla _a\nabla_b \gamma_{ce}(\lambda)|$ and the relevant weak--limits as $\lambda\searrow 0$.   This is not a minor technical point, since it is this non--uniformity in bounding the size of $\nabla _a\nabla_b \gamma_{ce}(\lambda)$  which is responsible for the sudden activation (at $\lambda = 0$) of the trace--less backreaction in GW. 

\noindent
{\em If we make $\nabla _a\nabla_b \gamma_{ce}(\lambda)$ uniformly bounded as $\lambda\searrow 0$, then there is no $\mu_{abcdef}$ to play with.}

\vspace{5pt}
\noindent
Summing up,  we have to require (as a necessary and sufficient condition for (\ref{weakassumption}) to hold and produce GW's results) that, (extending the GW enumeration of assumptions \emph{(i), (ii), (iii), (iv)}):
\\
\emph{(v)} For every given $\lambda\in (0, 1]$, there is a sequence 
$\{\nabla_b \gamma_{ce}^{(n)}(\lambda)\}_{n\in \mathbb{N}}$ 
of functions $\nabla_b \gamma_{ce}^{(n)}(\lambda)\in C^\infty(U)$ such that, as $ n \longrightarrow \infty$,
$\nabla_b \gamma_{ce}^{(n)}(\lambda)\longrightarrow \nabla_b \gamma_{ce}(\lambda)$ in $L^1_{\mathrm loc}(U)$, and
\begin{eqnarray}
(vi)\;\;\;\nabla _a\nabla_b \gamma_{ce}^{(n)}(\lambda)&\longrightarrow& F_{abce}(\lambda)\;,
\nonumber\\
(vii)\;\;\;\nabla _a\nabla_b \gamma_{ce}^{(n)}(\lambda)|_{\lambda\searrow 0}&\longrightarrow& 0\;, 
\end{eqnarray}
again in $L^1_{\mathrm loc}(U)$.  (These requirements simply follow from the standard characterization of  weak derivatives as limits of sequences of derivatives of smooth functions.)  Moreover,\\
\emph{(viii)} The sequence  $\nabla_d\nabla_b \gamma_{ce}^{(n)}(\lambda)\in C^\infty(U)\cap L^1_{\mathrm loc}(U)$ must not be uniformly bounded (as a function of $\lambda$) as $\lambda\searrow 0$.
If such a sequence exists, then we can identify the (weak) derivative $\nabla _a\nabla_b \gamma_{ce}^{(n)}(\lambda)$  of $\nabla_b \gamma_{ce}(\lambda)$ with $F_{abce}(\lambda)$ and this weak derivative will have the property $\wlim \nabla _a\nabla_b \gamma_{ce}^{(n)}(\lambda)\,=\,0$, required by the GW formalism.

These remarks show that $\wlim \nabla _a\nabla_b \gamma_{ce}(\lambda)\,=\,0$ is {\em not} a consequence of the GW hypotheses \emph{(ii), (iii), (iv)} but a strong {\em a priori} assumption that resembles GW's interpretation of the steel ball example (where they {\em a priori} kill the angular defects in the limit). To reach the conclusion that $\nabla _a\nabla_b \gamma_{ce}(\lambda)$ weakly vanishes as $\lambda\searrow 0$, and produces non--trivial results we must add to the GW assumptions 
$(ii)$--$(iv)$ the further hypotheses $(v)$--$(viii)$ described above, some of which lack physical justification. 

All in all, we cannot  conclude, with any reasonable level of geometrical and physical rigour, that  the weak--limit of  the (linear)  terms $g(0)\,\nabla \nabla \gamma(\lambda)$, present in $G_{ab}(\lambda)$, vanishes as $\lambda\searrow 0$.  A minimum requirement for this  to happen is a much stronger and unphysical control $(ii)$--$(viii)$ on the curve of metrics $\lambda\longmapsto g(\lambda)$ than that associated with GW's assumptions $(ii)$--$(iv)$. In particular, we wish to emphasize that the (implicit) non--uniformity of GW's weak ``averaging" is highly formal in a cosmological setting. GW's formalism depends on this assumption in an essential way. 

We conclude: replacing the non--uniform boundedness of $\nabla\nabla\,\gamma (\lambda)$, as $\lambda\searrow 0$, with a more natural boundedness assumption, 
{\em GW's formalism becomes empty}, insofar as, by removing the non--uniform boundedness (in
$\lambda$) requirement on $\nabla\nabla\gamma (\lambda )$ as $\lambda\searrow 0$, the {\em a
priori} assumptions \emph{(i)--(iv)} made by Green and Wald can only be
self--consistent if the tensor field $\mu_{abcdef} (0)$ vanishes identically.

\section{Comments on examples for backreaction that aim at including matter inhomogeneities}
\label{appendixC}

\subsection*{Example by Szybka et al.\/ }

Let us consider Szybka et al.'s \cite{szybka} example in
relation to GW's claim that backreaction is trace--free. The example is based on the
Wainwright--Marshman metric:
\begin{equation}
{\mathrm d}s^2 = t^{2m}e^n\left(-{\mathrm d}t^2 + {\mathrm
  d}z^2\right) + t^{1/2}\left[{\mathrm
  d}x^2+\left(t+w^2\right){\mathrm d}y^2 + 2w\, {\mathrm d}x \mathrm{d}y\right]
\;,
\end{equation}
where $t > 0 ,\;-\infty<x,y,z,<+\infty$, 
$m$ is a free parameter, $n$ and $w$ are functions of a single variable $u: = t-z$, 
and the field equation
\begin{equation}
  \label{appcefe}
  \mathrm{d}n/\mathrm{d}u = (\mathrm{d}w/\mathrm{d}u)^2
\end{equation}
helps to simplify the Einstein tensor.
The stress--energy tensor is that of a perfect fluid with equation of state
\begin{equation}
\varrho = p = \frac{1}{8\pi}(m + 3/16)t^{-2(m+1)}e^{-n}\;,
\end{equation}
so the weak energy condition holds for $m \geq -3/16$.  
Szybka et al.\/ \cite{szybka} set $w := \lambda \sin(u/\lambda)$
which gives  $n = \frac{1}{2}\left(u + \frac{1}{2}\lambda \sin(2u/\lambda)\right) $. In the limit: $\liml w = 0$ and $\liml n = \frac{1}{2}u$, hence (\ref{appcefe}) no longer holds for the background.

The Ricci scalar of the metric $g_{ab}(\lambda)$ is
$R(\lambda) = -\frac{1}{8}\left(16m+3\right)t^{-2(m+1)}e^{-n(\lambda)}$,
where for the background metric $g_{ab}(0)$, 
$n$ should be substituted by $\liml n$. 
This can be used as follows to show
that, although the derivatives of metric deviations and quadratic products
thereof in this example are singular in the limit as $\lambda$ goes to zero, the Ricci scalar
and the stress--energy tensor are not and thus cannot produce backreaction from inhomogeneities. 
First, let us subtract the Einstein equations for $\lambda > 0$, 
\begin{equation}
\label{C4}
R_{ab}(g(\lambda)) - \frac{1}{2}g_{ab}(\lambda)R(g(\lambda)) = 8\pi T_{ab}(\lambda)\;,
\end{equation}
from the background dynamical equation; for $\lambda =0$ (using the notation $t_{ab}(0)$, {\em cf.} Sect.~\ref{nobackreaction}), we have:
\begin{equation}
\label{C5}
R_{ab}(g(0)) - \frac{1}{2}g_{ab}(0)R(g(0)) = 8\pi \left(T_{ab}(0)+t_{ab}(0)\right) \; .
\end{equation}
We  take the {$\wlim$} of the difference (\ref{C5})--(\ref{C4})  (recall that $t_{ab}(0)$ remains unaffected by the weak--limit operator):
\begin{equation}
t_{ab}(0) = \wlim_{\lambda\searrow 0} \left(R_{ab}(g(0))-R_{ab}(g(\lambda))\right),
\end{equation}
because $\wlim \left(R(g(0))-R(g(\lambda))\right) = 0\;$,
$\wlim \left(g_{ab}(0) R(g(0))-g_{ab}(\lambda) R(g(\lambda))\right) = 0$,
and $\wlim \left(T_{ab}(0)-T_{ab}(\lambda)\right) = 0.$

Introducing the trace--free Ricci tensor $S_{ab} = R_{ab} - \frac{1}{4}g_{ab} R$,
we get:
\begin{equation}
t_{ab}(0) = \wlim_{\lambda\searrow 0} \left(S_{ab}(g(0))-S_{ab}(g(\lambda))\right)\;,
\end{equation}
since again $\wlim_{\lambda\searrow 0} \left(g_{ab}(0) R(g(0))-g_{ab}(\lambda) R(g(\lambda))\right) = 0$. 
From this we see that the backreaction term $t_{ab}$ entirely emerges from the  trace--less $\lambda$--dependent curvature ($S_{ab}(g(\lambda))$). The Ricci scalar and the stress--energy tensor, where the density inhomogeneities are encoded, cancel out. Since the Wainwright--Marshman spacetimes are interpreted as cosmological models with gravitational waves, we can attribute all the backreaction present in this example to gravitational radiation, not to density inhomogeneities.

\subsection*{Example by Green and Wald}

Now we consider the example provided by GW in \cite{gw3} (section 4).
We recall GW's assumption {\em (i)}\cite{gw1}, {\em  i.e.},
that there exists a family of metrics $g_{ab}(\lambda)$ and a smooth function $C_1$ such that:
\begin{itemize}
\item $g_{ab}(1)$ represents the real Universe, while $g_{ab}(0)$ is the background, averaged metric;
\item $G_{ab}(g(\lambda)) = 8\pi T_{ab}(\lambda)$ for all $\lambda>0$
  and $T_{ab}(\lambda)$ obeys the weak energy condition; and
\item $h_{ab}(\lambda) = |g_{ab}(0) - g_{ab}(\lambda)| < \lambda C_1\;.$
\end{itemize}
What does it imply for $g_{ab}(\lambda)$ to be a solution of the Einstein
equations? For $\lambda =1$ this is true by definition, since we assume that
the real Universe is well--described by general
relativity. However, while we have a straightforward prescription of how
to construct $G_{ab}(g(\lambda))$ for $1>\lambda>0$, the situation
with $T_{ab}(\lambda)$ is not so clear.

We can distinguish two reasonable approaches:
\begin{enumerate}
\item we calculate $G_{ab}(g(\lambda))$ and {\em define}
  $T_{ab}(\lambda) := \frac{1}{8\pi} G_{ab}(\lambda)$
  (this is the option chosen by Green and Wald); or \label{jeden}
\item we calculate $T_{ab}$ for a given $\lambda$ from its definition in terms 
of $\mathcal{L}_{\mathrm{m}}$, the non-gravitational part of 
the Lagrangian density of the Einstein--Hilbert action for matter, and its 
functional derivative, i.e.,
\begin{equation}
  T_{ab} :=
  2\;\frac{\delta \mathcal{L}_{\mathrm{m}}}{\delta g_{ab}} +
  g^{ab}\;\mathcal{L}_{\mathrm{m}}\;. 
  \label{e-Lag-defn-Tab}
\end{equation}
\label{dwa}
\end{enumerate} 
\noindent
To see that these two possibilities are, in general, different, let
us consider Green and Wald's example for both of them.
By assumption, there is an FLRW background metric $g_{ab}(0)$ and a conformally related family of metrics:
\begin{eqnarray}
\fl \qquad
  g_{ab}(\lambda) = \Omega^2(\lambda)g_{ab}(0)\;,\quad{\rm where}\quad
  \ln \Omega(\lambda) = \lambda A \left(\sin \frac{x}{\lambda} + \sin
  \frac{y}{\lambda} + \sin \frac{z}{\lambda} \right).
  \label{eq-defn-Omega}
\end{eqnarray}
The Einstein tensor built out of $g_{ab}(\lambda)$, i.e., $G_{ab}(g(\lambda))$, 
is related to the FLRW Einstein tensor $G_{ab}(g(0))$ by a purely geometrical formula:
\begin{eqnarray}
    \label{eq2}
    \fl
   \qquad\qquad G_{ab}(g(\lambda)) =& G_{ab}(g(0))- \left(2\nabla_a\nabla_b \ln
    \Omega - 2 g_{ab}(0)g^{cd} (0) \nabla_c \nabla_d \ln \Omega
    \nonumber \right.\\
    &\left. - 2 (\nabla_a \ln \Omega)(\nabla_b \ln \Omega) - g_{ab} (0)
    g^{cd} (0) (\nabla_c \ln \Omega)(\nabla_d \ln \Omega)\right).
\end{eqnarray}
Green and Wald show (see (4.4) in \cite{gw3}) that the effective stress--energy tensor in the weak--limit reads:
\begin{eqnarray}
  \label{eq1}
\fl \qquad\qquad  t_{ab} (0) &=& \frac{1}{8\pi}G_{ab}(g (0)) - T_{ab}(0) 
  =\frac{1}{8\pi} \wlim_{\lambda\searrow 0} \left( G_{ab}(g (0)) - G_{ab}(g(\lambda))\right),
\end{eqnarray}
and that the backreaction found by the GW prescription has a trace corresponding to a $P = -\frac{5}{3}\rho$ fluid.

Now we have to choose in which way we define the  stress--energy tensor for $0<\lambda<1$. \\ 
On the one hand from (\ref{jeden}) :
\begin{eqnarray}
  \label{e-Tab-assume-traceless}
8\pi T_{ab}(\lambda) =& G_{ab}(g(\lambda))\;,
  \end{eqnarray}
   into which (\ref{eq2}) has to be substituted, while on the other hand  from (\ref{dwa}):
\begin{equation}
  \label{eq4}
  T_{ab}(\lambda) = \Omega^{-2}T_{ab}(0)\;.
\end{equation}
The reason for the transformation rule (\ref{eq4}) is conformal invariance of the matter 
action:
\begin{equation}
  \label{e-action-S-lambda}
  S(\lambda) = 
  \int \mathcal{L}_{\mathrm{m}}(\lambda) \sqrt{-g(\lambda)}\mathrm{d}^4x = 
  \int \mathcal{L}_{\mathrm{m}}(0)  \sqrt{-g(0)}\mathrm{d}^4x = S(0)\;,
\end{equation}
where the Lagrangian densities transform according to
\begin{equation}
  \mathcal{L}_{\mathrm{m}}(\lambda) = \Omega^{-4}\mathcal{L}_{\mathrm{m}}(0)\;,
\end{equation}
so that (\ref{e-Lag-defn-Tab}) gives
\begin{eqnarray}
  \label{e-Tab-Lag-action-check}
  T^{ab}(\lambda) &=&
  \frac{2}{\sqrt{-g(\lambda)}}\;\;\frac{\delta}{\delta
    g_{ab}(\lambda)}\left(\sqrt{-g(\lambda)}\mathcal{L}_{\mathrm{m}}(\lambda)\right)
  \nonumber \\
  &=& \Omega^{-4}\frac{2}{\sqrt{-g(0)}} \;\; \frac{\partial
    g_{cd}(0)}{\partial g_{ab}(\lambda)} \;\; \frac{\delta}{\delta
    g_{cd}(0)}\left(\sqrt{-g(0)}\mathcal{L}_{\mathrm{m}}(0)\right);
\end{eqnarray}
lowering of indices using $g_{ae}(\lambda)$ and $g_{bf}(\lambda)$ gives (\ref{eq4}) (see, e.g., \cite{Dabrowski}).
In (\ref{e-action-S-lambda}) and (\ref{e-Tab-Lag-action-check}), a non--subscripted $g$
indicates the determinant of the metric.
 
To show that case (\ref{jeden}) is in general incompatible with case (\ref{dwa}) we proceed as follows.
Let us assume that for $\lambda = 1$,
the metric and stress--energy tensor represent the `real' Universe or, as in the case presented, the toy universe model that we wish to average.
Define $ f := \Omega(\lambda)\Omega^{-1}(\lambda = 1)$.

We have for case (\ref{jeden}):
\begin{eqnarray}
\fl
  G_{ab}(g(1)) = 8\pi T_{ab}(1) \ ;\ 
  g_{ab}(1) = \Omega^2(\lambda = 1)g_{ab}(0) \ ; \ 
    g_{ab}(\lambda) = \Omega^2(\lambda)g_{ab}(0) \;,\nonumber
\end{eqnarray}
which implies that 
\begin{eqnarray}
  g_{ab}(\lambda) = \Omega^2(\lambda)\Omega^{-2}(\lambda = 1)g_{ab}(1)\; .
  \label{eq-gablambda-gab1}
\end{eqnarray}
Thus, $g_{ab}(\lambda<1)$ and $g_{ab}(1)$ are conformally related via 
$g_{ab}(\lambda) = f^2 g_{ab}(1)$, 
since $f$ is a ratio of exponentials of smooth real--valued functions
(see (\ref{eq-defn-Omega})), and thus smooth and real--valued.

For case (\ref{dwa}):
\begin{eqnarray}
  \fl \qquad\qquad\qquad T_{ab}(\lambda) &= &f^{-2}T_{ab}(1)  \label{eqw}\\
  G_{ab}(g(\lambda)) &=& G_{ab}(g(1))-
  \left(2\widetilde{\nabla}_a\widetilde{\nabla}_b \ln f - 2
  g_{ab}(1)g^{cd} (1) \widetilde{\nabla}_c \widetilde{\nabla}_d \ln f
  \nonumber- \right. \nonumber \\
\label{eq12}  
&&\left.  2 (\widetilde{\nabla}_a \ln f)(\widetilde{\nabla}_b \ln f)
  - g_{ab} (1) g^{cd} (1) (\widetilde{\nabla}_c \ln
  f)(\widetilde{\nabla}_d \ln f)\right),
\end{eqnarray}
where the covariant derivative $\widetilde{\nabla}$ is compatible with the metric $g_{ab}(1)$ and the geometrical formula (\ref{eq12}) is obtained directly from (\ref{eq2}),
and thus is also valid in case (\ref{dwa}).
Combining (\ref{eqw}) and (\ref{eq12}) and imposing the Einstein equations for $\lambda > 0$ we 
obtain:
\begin{eqnarray}
8\pi T_{ab}(1) &=& \frac{1}{(1-f^{-2})}
  \left(2\widetilde{\nabla}_a\widetilde{\nabla}_b \ln f - 2
  g_{ab}(1)g^{cd} (1) \widetilde{\nabla}_c \widetilde{\nabla}_d \ln f
  \nonumber- \right. \nonumber \\
  &&\left.  2 (\widetilde{\nabla}_a \ln f)(\widetilde{\nabla}_b \ln f) -
  g_{ab} (1) g^{cd} (1) (\widetilde{\nabla}_c \ln
  f)(\widetilde{\nabla}_d \ln f)\right),
\end{eqnarray}
which is a contradiction since the left--hand side is fixed and the right--hand side depends on
$\lambda$ (lengthy but easy calculations show that the dependence on $\lambda$ does not cancel).

We conclude that, within the GW formalism (as their
explicit example shows), it is not in general possible to build a
family of metric--dependent tensors using the method (\ref{dwa}) that
obeys the Einstein equations. This gives us a strong indication of the nature of the $T_{ab}(\lambda)$ family obtained with method (\ref{jeden}).

To see this explicitly let us look at GW's choice (\ref{jeden}): using the purely geometrical relation (\ref{eq12}) and the Einstein equations for $\lambda >0$, we have\footnote[23]{In the arXiv preprint version~1 of this paper we erroneously referred to Eq. (\ref{eqw}), as has been kindly pointed out in \cite{gwnote}.}:
\begin{eqnarray}
  8\pi T_{ab}(\lambda) =& 8\pi T_{ab}(1)-
  \left(2\widetilde{\nabla}_a\widetilde{\nabla}_b \ln f - 2
  g_{ab}(1)g^{cd} (1) \widetilde{\nabla}_c \widetilde{\nabla}_d \ln f
  \nonumber- \right. \nonumber \\
  &\left.  2 (\widetilde{\nabla}_a \ln f)(\widetilde{\nabla}_b \ln f) -
  g_{ab} (1) g^{cd} (1) (\widetilde{\nabla}_c \ln
  f)(\widetilde{\nabla}_d \ln f)\right).&
\end{eqnarray}
Now we take the weak--limit of both sides.

The weak--limit of $T_{ab}(g(1))$ is (where $f^{ab}$ is a test tensor field):
\begin{equation}
  \wlim_{\lambda\searrow 0} T_{ab}(g(1)) = \int \mathrm{d}^4x \sqrt{-g(0)} T_{ab}(g(1))f^{ab}  \;,
\end{equation}
which we can naturally associate with an averaged stress--energy
tensor usually denoted $\langle T_{ab} \rangle$ and denoted
$T_{ab}(0)$ by GW (according to GW: ``we may interpret $T_{ab}(0)$ as representing the matter stress--energy tensor averaged over small scale inhomogeneities'', \cite{gw1} p. 13). Since the weak--limit of the remaining terms on the right--hand side does not disappear, we see that $\wlim T_{ab}(\lambda) \neq T_{ab}(0)$. As a consequence, in a more realistic case
we should not expect $T_{ab}(0)$ to match the FLRW stress--energy tensor obtained by averaging the density inhomogeneities. It seems that the $T_{ab}(\lambda)$
defined via the Einstein field equations already contains a backreaction effect
and thus a part of backreaction is absorbed into the definition of $T_{ab}(0)$. In other words, forcing the existence of stress--energy tensors that obey
the Einstein equations makes the weak--limit of this family of tensors
distinct from the averaged stress--energy tensor.   

In summary, we are not aware of an example of a metric family satisfying
the GW conditions that satisfactorily describes backreaction from
matter inhomogeneities.

\clearpage
\section*{References}
\bibliographystyle{iopart-num}
\providecommand{\newblock}{}

\end{document}